%% file: ms_v0.tex
\documentclass[twocolumn]{aastex61}
\usepackage{natbib}
\usepackage{amsmath}
\newcommand{\ssst}{\scriptscriptstyle}
\newcommand{\E}[1]{\times 10^{#1}}

\newcommand{\RA}[3]{{#1}^{{\rm h}}{#2}^{{\rm m}}{#3}^{{\rm s}}}
\newcommand{\decl}[3]{{#1}^{\circ}{#2}'{#3}''}

\newcommand{\ps}{\,{\rm s}^{-1}}
\newcommand{\yr}{\,{\rm yr}}    
\newcommand{\Msun}{M_{\odot}}
\newcommand{\cm}{\,{\rm cm}}    
\newcommand{\km}{\,{\rm km}}
\newcommand{\kms}{$\km\ps$}

\newcommand{\pc}{\,{\rm pc}}
\newcommand{\erg}{\,{\rm erg}}  
\newcommand{\K}{\,{\rm K}}

\newcommand{\NH}{N_{\ssst\rm H}}

\newcommand{\Tmb}{T_{\rm mb}}
\newcommand{\Tk}{T_{\rm k}}

\newcommand{\nH}{n_{\ssst\rm H}}  

\newcommand{\nHH}{n({\rm H}_{2})} 
\newcommand{\NHH}{N({\rm H}_{2})}
\newcommand{\VLSR}{V_{\ssst\rm LSR}}

\newcommand{\du}{d_{3.4}}

\newcommand{\snr}{Cas~A}
\newcommand{\twCO}{$^{12}$CO}   
\newcommand{\thCO}{$^{13}$CO}
\newcommand{\HCOp}{HCO$^+$}   
\newcommand{\HH}{H$_2$}       

\newcommand{\Jotz}{$J$=1--0}    
\newcommand{\Jtto}{$J$=2--1}

\accepted{August 7, 2018}
\submitjournal{ApJ}

\shorttitle{Molecular gas toward supernova remnant Cassiopeia A}
\shortauthors{Zhou et al.}

\begin{document}

\title{Molecular gas toward supernova remnant Cassiopeia~A}

\correspondingauthor{Ping Zhou}
\email{p.zhou@uva.nl}

\author{Ping Zhou}
\affil{Anton Pannekoek Institute for Astronomy, University of Amsterdam, Science Park 904, 1098 XH Amsterdam, The Netherlands}
\affil{School of Astronomy and Space Science, Nanjing University,
163 Xianlin Avenue, Nanjing, 210023, China}

\author{Jiang-Tao Li}
\affil{Department of Astronomy, University of Michigan, 311 West Hall, 1085 S. University Ave, Ann Arbor, MI, 48109-1107, U.S.A.}

\author{Zhi-Yu Zhang}
\affil{Institute for Astronomy, University of Edinburgh, Royal Observatory, Blackford Hill, Edinburgh EH9 3HJ, UK}
\affil{ESO, Karl Schwarzschild Strasse 2, D-85748 Garching, Munich, Germany}

\author{Jacco Vink}
\affil{Anton Pannekoek Institute for Astronomy, University of Amsterdam, Science Park 904, 1098 XH Amsterdam, The Netherlands}
\affil{GRAPPA, University of Amsterdam, Science Park 904, 1098 XH Amsterdam, The Netherlands}
\affil{SRON, Netherlands Institute for Space Research, Sorbonnelaan 2, 3584 CA Utrecht, The Netherlands}

\author{Yang Chen}
\affil{School of Astronomy and Space Science, Nanjing University,
163 Xianlin Avenue, Nanjing, 210023, China}
\affil{Key Laboratory of Modern Astronomy and Astrophysics, Nanjing University, Ministry of Education, China}

\author{Maria Arias}
\affil{Anton Pannekoek Institute for Astronomy, University of Amsterdam, Science Park 904, 1098 XH Amsterdam, The Netherlands}

\author{Daniel Patnaude}
\affil{Smithsonian Astrophysical Observatory, Cambridge, MA 02138, USA}

\author{Joel N.\ Bregman}
\affil{Department of Astronomy, University of Michigan, 311 West Hall, 1085 S. University Ave, Ann Arbor, MI, 48109-1107, U.S.A.}



\begin{abstract}

We mapped \twCO~\Jotz, \twCO~\Jtto, \thCO~\Jotz, and \thCO~\Jtto\ lines 
toward supernova remnant (SNR)
Cassiopeia~A with the IRAM~30m telescope.
The molecular clouds (MCs) along the line of sight of \snr\ do not show optically 
thin, shock-broadened \twCO\ lines ($\Delta V\le 7\km\ps$ toward \snr), or 
high-temperature features  from shock heating ($\Tk\le 22~\K$ toward
\snr).
Therefore, we suggest that there is no physical evidence
to support that the SNR is impacting the molecular gas.
All the detected MCs are likely in front of \snr, as implied by the 
\HCOp\ absorption line detected in the same velocity ranges.
These MCs contribute H$_2$ column densities of $5\E{21}~\cm^{-2}$,
$5\E{21}~\cm^{-2}$, and $2\E{21}~\cm^{-2}$ in the west, south, 
and center of the SNR, respectively.
The 20~K warm gas at $\VLSR\sim -47~\km\ps$ is distributed along a large-scale molecular ridge in the south of Cas~A.
Part of the gas is projected onto Cas~A, providing a foreground 
H$_2$ mass of $\sim 200 (d/3~{\rm kpc})^2~\Msun$, consistent with 
the mass of cold dust  (15--20~K; 2--4~$\Msun$) found in front 
of the SNR.
We suggest that the 20~K warm gas is heated by background cosmic rays 
with an ionization rate of $\zeta({\rm H_2})\sim 2\E{-16} \ps$.
The cosmic rays and X-ray emission from \snr\ are excluded as the 
heating sources of the clouds.

\end{abstract}




\section{Introduction} \label{sec:intro}

Cassiopeia~A, a.k.a., Cas~A  (G$111.7-2.1$), is the remnant of a 
young supernova that exploded around AD $1681\pm 19$ \citep{fesen06}
at a distance of  $3.4^{+0.3}_{-0.1}$~kpc \citep{reed95}.
Due to its youth and brightness, \snr\ is among the best-observed 
supernova remnants(SNRs) and the prototype for many aspects of SNR studies. 
The current manifestations of an SNR are the combined results
of the intrinsic supernova properties and the environment 
it evolves in.
The study of the environment not only provides crucial information
of the SNR itself,  but also helps us to understand
how supernova explosions affect the interstellar medium of the galaxies.
In spite of a wealth of observations of \snr\ across the electromagnetic 
wavelengths, there is no consensus on the immediate SNR environment.

It is a common view that \snr\ evolves in the stellar winds of the 
progenitor star.
The early suggestion of the idea was based on the detections of the 
quasi-stationary flocculi \citep{vandenbergh73}.
The circumstellar medium of \snr\ was likely produced by slow and dense 
winds, characterized by a 
$\rho_{\rm w}\propto r^{-2}$  density profile \citep[e.g.,][]{vink04a}.
This picture well explains the positions of the SNR's 
forward and reverse shocks \citep{chevalier03},
the X-ray properties of shocked ejecta knots 
\citep{laming03,hwang09}, and the characteristics of the shocked 
ambient gas \citep{lee14}. 
The preshock density estimated from the X-ray analysis is
around $0.9\pm0.3~\cm^{-3}$,  without strong variation 
at different position angles \citep{lee14}.

Outside the SNR boundary the environmental gas is cooler and its
relation with \snr\ less clear.
The difficulty is how to distinguish the immediate 
environment and the interstellar medium along the line of sight.
The dense gas along the line of sight of \snr\ is mainly distributed
in the velocity range of $-50$ to $-30\km\ps$ (associated with
the Perseus arm), and at $\sim 0~\km\ps$ (associated the Orion
spur), as observed in molecular emission and absorption lines
\citep{dejager78,batrla83,batrla84,goss84,bieging86,wilson93,anantharamaiah94,reynoso02,kilpatrick14},
\ion{H}{1} absorption \citep{mebold75,bieging91,schwarz97},
[\ion{C}{1}] \citep{mookerjea06},
and carbon recombination lines \citep{payne89,anantharamaiah94,kantharia98, mookerjea06,oonk17, salas17,salas18}.
\citet{reynoso97b} found some small absorption \ion{H}{1} features
at the lower LSR velocity of $-69$ to $-62\km\ps$, and suggested that
they are from neutral knots driven away by the progenitor wind of \snr.
If this velocity of foreground \ion{H}{1} structures represents the upper 
limit of \snr's systemic velocity, it may indicate that \snr\ is 
not associated with any of the clouds in the velocity range of $-50$ to $-30~\km\ps$.
This idea is supported by the carbon recombination line 
studies, which suggest that the gas at 
$\VLSR\lesssim -47\km\ps$ is at least 100~pc 
in front of \snr\ \citep{kantharia98, salas17}. 
However, since the molecular gas is traced by molecular lines, 
the relationship between the molecular gas and \snr\ should 
be studied with molecular observations.

Warm molecular gas with a kinetic temperature of 20~K
was found at $\VLSR\sim -47\km\ps$ \citep{wilson93}, using
\thCO\ and \twCO\ observations with IRAM 30~m, while
the typical temperature of an interstellar giant molecular cloud (MC) is 
$\sim 10~\K$.
The association between \snr\ and molecular clouds (MCs)
was proposed in a recent study using the SMT 12~m telescope
\citep{kilpatrick14},
based on a subtle line broadening (line width of 
$\sim 6\km\ps$) of the \twCO\ emission at $\VLSR\sim -37\km\ps$
and $\sim -47 \km\ps$ in the west and south of the SNR.
If true, the MCs could provide dense targets for the shock and 
cosmic-ray (CR) protons to interact, making the properties of
the MCs important constraints for the studies of the SNR's shock
and CRs.
However, an association between \snr\ and MC is inconsistent
with the low preshock density found in X-rays \citep{lee14}
and the suggestion that the clouds at $\VLSR=-50$--$0\km\ps$ are all foreground gas.

Motivated by the aforementioned problems,
we performed new molecular line observations toward \snr\ and
its environment, aiming to answer the following questions: Is the SNR interacting
with MCs? what are the properties of the MCs? 
We also examine what is the heating source of the 20~K MC 
at $-47\km\ps$.
Our molecular mapping observations have the best 
angular resolution of these lines to date for this SNR
($11''$ at \twCO\ \Jtto), 
and good sky coverage (83 arcmin$^2$ at \twCO\ \Jtto).

\section{Data} \label{sec:data}
\subsection{IRAM 30~m observation}

We performed 3 and 1\,mm heterodyne observations toward \snr\ with the
IRAM 30~m telescope, during 2016 July 12, September 24, 
November 15--19, and December 5 (Project IDs 145-15, 053-16, and 136-16; PI: Jiang-Tao Li),
and 2017 September 16--18 (029-17; PI: Ping Zhou).
The observations consist of mappings of the SNR and line surveys of a few
positions of interest.
This work focuses on the \twCO\ and \thCO\ mapping observations, 
but also uses the \HCOp\ observation at one position.

\begin{deluxetable*}{lcccccc}
\tablecaption{Observational information \label{tab:obs}}
\tablehead{
\colhead{Line} & \colhead{Frequency} & 
{Instrument}  &  HPBW & \colhead{Map Area} & 
\colhead{Average/Central rms\tablenotemark{a}} \\
& (GHz) & & (arcsec) & (arcmin$^2)$ & (K)
}
\startdata
\twCO~\Jtto & 230.540  & HERA+EMIR & 10.7 & 83 & 0.65/0.29 \\
\thCO~\Jtto & 220.399  & HERA+EMIR & 11.2 & 54 & 0.30/0.21 \\
\twCO~\Jotz & 115.271  & EMIR      & 21.3 & 79 & 0.27/0.22 \\
\thCO~\Jotz & 110.201  & EMIR      & 22.3 & 69 & 0.12/0.09 \\
\HCOp~\Jotz & 89.188   & EMIR      & 27.6 & -  & 0.008\\
\enddata
\tablenotetext{a}{The rms of the spectrum in each pixel 
($\approx 1/2$ of the HPBW) with a velocity resolution of 0.5~\kms\ in the velocity range from $-100\km\ps$ to
$50~\km\ps$. 
The central rms is
        calculated from the interior  $7'\times 7'$ region centered at
        ($\RA{23}{23}{27}$, $\decl{58}{48}{40}$, J2000), where deeper
observations were taken. } 
\end{deluxetable*}

We mapped \twCO~\Jotz\ (at 115.271 GHz), \thCO~\Jotz
(at 110.201~GHz), \twCO~\Jtto\ (at 230.540~GHz) and \thCO~\Jtto\ (at
220.399~GHz) emission toward \snr\ and its vicinity, using the on-the-fly
mode with the 9-pixel dual-polarization HEterodyne Receiver Array
(HERA) and the Eight MIxer Receiver (EMIR).  On 2016 July 12 and 
September 24, 
the HERA was used to map the  \twCO~\Jtto\ and \thCO~\Jtto\ lines.
The back end of fast Fourier transform spectrometers (FTSs) provided a frequency 
resolution of 50~kHz, corresponding to a velocity
resolution of 0.063~\kms. 
Bad spectra produced by problematic pixels of HERA were removed. 
For the rest of the mapping observation, we used EMIR to
observe the \twCO\ and \thCO\ lines  at 1 and 3~mm bands, 
simultaneously. 
In the EMIR observations FTSs were tuned to the 
200~kHz resolution, which 
provided a velocity resolution of 0.25~\kms\ for the \twCO~\Jtto\ line and 0.5~\kms\ for the \twCO~\Jotz\ line.  
Although we optimized the observations for the velocity
range of $\VLSR=-100$-- $50\km\ps$, the FTS units covered 
a wide frequency range of 8~GHz at 200~kHz resolution, 
corresponding to a velocity coverage of $\sim 10^4 \km\ps$ for the
\twCO~\Jtto\ line and $\sim 2\E{4}\km\ps$ for the \twCO~\Jotz\ line.
By inspecting the spectra in the wide velocity
range, we only found strong emission 
lines at the velocity range from $\VLSR\sim-50\km\ps$ to $\VLSR\sim 0~\km\ps$.
Given that we aim to study the interstellar MCs,
we only provide detailed analysis of the data
in the velocity range from $-100$ to $50~\km\ps$, where the
baselines were reduced with linear lines.
We do not show the results of the wide velocity range, also 
because the baselines of the very wide band 
are not linear or stable in different observational epochs,  
but sometimes contaminated by instrumental features and large 
noises at some velocities.
The half-power beam width (HPBW) of the telescope was $10\farcs{7}$ at 
230~GHz and $21\farcs{3}$ at 115~GHz. 
The main-beam efficiencies at these two frequencies were 59\% and 78\%,
respectively.

We also carried out a 1~mm line survey toward a few positions in and near
\snr, using the positional switch mode.  As this paper focuses on
the relation of \snr\ to its environment, we only use one spectrum of 
HCO$^+$ \Jotz\ at the western radio peak
($\RA{23}{23}{10}$, $\decl{58}{48}{40}$, J2000), where absorption lines are seen 
and will provide clues to the location of MCs. 
The rest of the survey data will be analyzed in another paper.
The back-end FTS was working at the frequency resolution of 50~kHz, providing
a velocity resolution of 0.17~\kms\ at a frequency of 89.188~GHz.
The beam efficiency was 81~\%  at this frequency.

The detailed observation information used in this work is tabulated in Table~\ref{tab:obs}.
The IRAM~30m data were reduced with the
GILDAS software\footnote{http://www.iram.fr/IRAMFR/GILDAS/}.

\subsection{FCRAO archival data}

We retrieve large-scale mapping data of \twCO~\Jotz\ produced 
by the Canadian Galactic Plane Survey (CGPS).
The original data are from ``The Five College Radio Astronomy Observatory (FCRAO) CO Survey of the Outer Galaxy''
\citep{heyer98}.
The data have been regridded from the FCRAO $50\farcs{22}$ pixel  scale to the CGPS $18''$ 
pixel scale, with a velocity step of 
0.82 \kms. Prior to spatial regridding 
the data were smoothed to 
the Nyquist resolution limit of $1\farcm{674}$.
A main-beam efficiency of 0.48 is adopted.

\section{Results} \label{sec:results}
\subsection{Distribution of the MCs}

\begin{figure*}
\epsscale{1}
\plotone{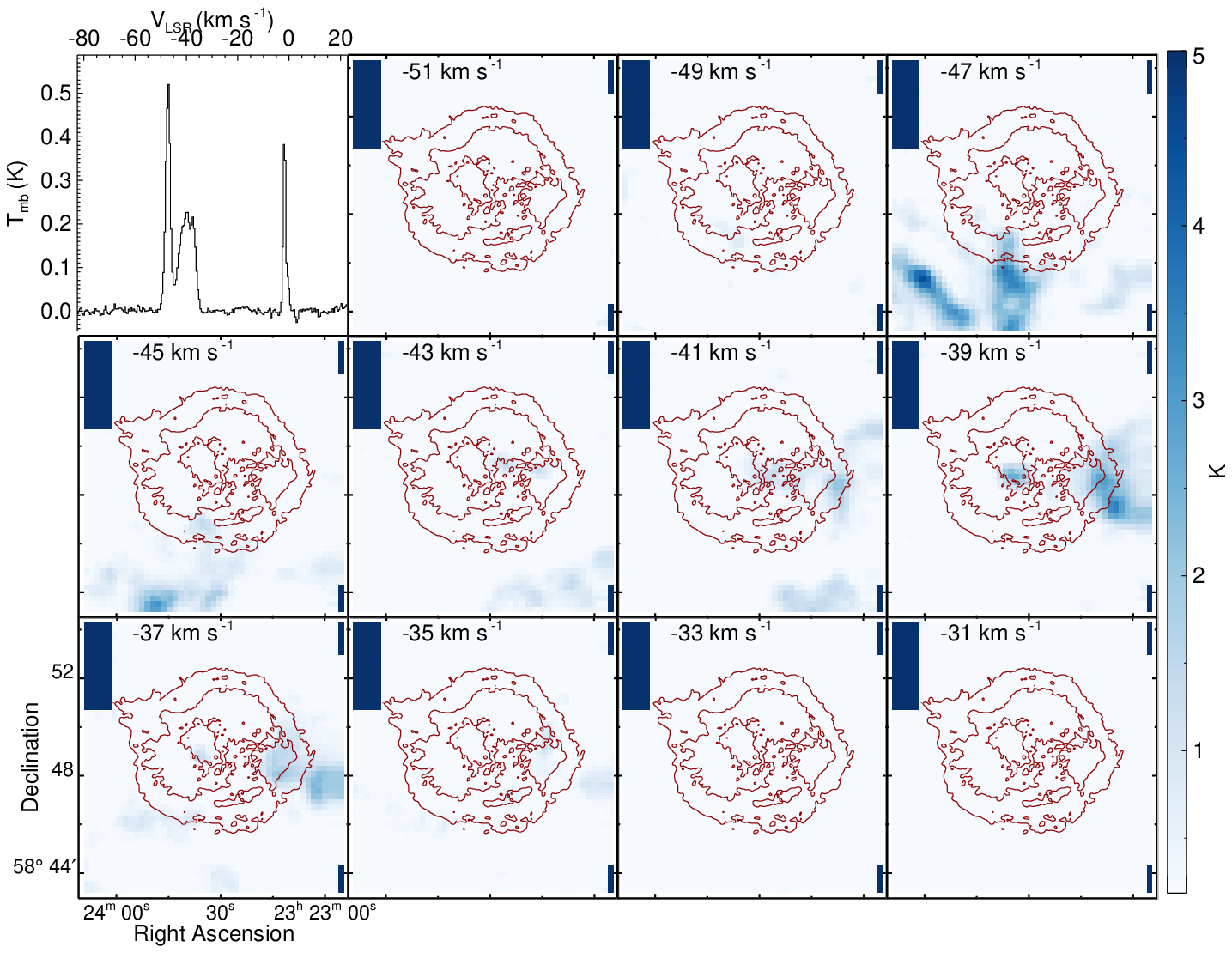}
\caption{
Channel map of the main-beam temperature of \thCO~\Jotz\ emission, 
overlaid with contours of {\it Chandra} X-ray emission. 
The same contours are used for the maps in the rest of the paper.
The map size is $0\fdg{14} \times 0\fdg{15}$, with a pixel size of $11''$.
The top-left panel shows the spectrum averaged over the FOV.
The corners in saturated blue color were not covered by the observation.
\label{fig:13co10_grid}}
\end{figure*}

\begin{figure*}
\epsscale{1}
\plotone{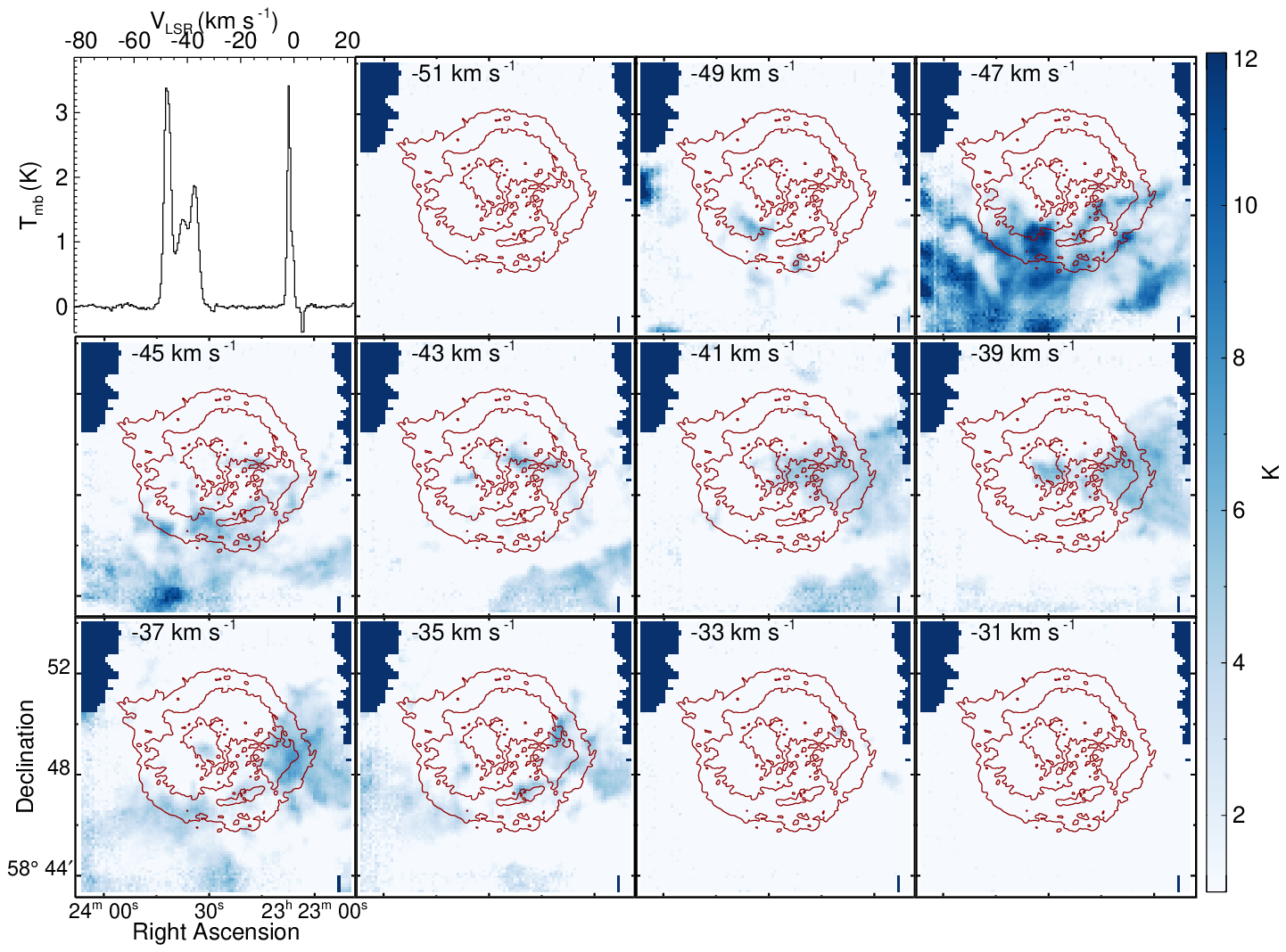}
\caption{
Channel maps of the main-beam temperature of \twCO~\Jtto. 
The map size is $0\fdg{15} \times 0\fdg{15}$, with a pixel size of $5\farcs{5}$.
The top-left panel shows the spectrum averaged over the FOV.
\label{fig:12co21_grid}}
\end{figure*}

The \thCO\ and \twCO\ emission is shown  at the local standard 
of rest velocities $\VLSR=-50$ to $-30\km\ps$, and $-2$\kms, respectively,
as indicated by spectra averaged in the field of view (FOV; see the top-left panels of Figure~\ref{fig:13co10_grid} and \ref{fig:12co21_grid}). 
We only analyze the molecular gas at the velocity range
between $-50\km\ps$ and $-30\km\ps$, since the gas at $\VLSR\sim 0 \km\ps$ 
corresponds to a local component \citep{dame01} that is unrelated to the remnant.

The distribution of MC column density can be described with the
\thCO\ \Jotz\ map, 
quiescent  molecular gas, given that \thCO\ emission is found to be optically
thin ($\tau_{\rm ^{13}CO}<1$) for the MCs toward \snr.
Figure~\ref{fig:13co10_grid} shows the \thCO\ distribution at different
velocities.
The molecular gas is mainly distributed in the south 
($\VLSR=-47\km\ps$)  and west ($\VLSR=-41$ to $-35\km\ps$),
with some clouds projected onto the SNR.
The north of the SNR does not show any MC emission, as also 
the \thCO~\Jtto\ data show a similar distribution to the
\thCO~\Jotz\ data.

The \twCO~\Jtto\ channel images reveal more extended structures from 
both high and low column density parts of the MCs  (see Figure~\ref{fig:12co21_grid}). 
At $-47\km\ps$, some bright filamentary structures are shown near the SNR,
which are not clearly shown in other velocity channels, indicating a small 
velocity dispersion of these structures.
The western clouds are fainter but span a larger velocity range of 
$-41~\km\ps$ to $-35\km\ps$.
Although the northern part of SNR shows very little molecular gas, 
there is a faint, thin filament outside the SNR northeastern boundary 
($-37\km\ps$) and  a small cloud to the north ($-41\km\ps$).

\subsection{Temperatures of the MCs} \label{sec:temp}

The measured main-beam temperature of the CO emission 
is determined by the excitation 
temperature $T_{\rm ex}$,  optical depth $\tau$, and beam filling factor $f$
($f<1$ if the clumpy gas is unresolved with the telescope's beam):

\begin{equation} \label{equ:tmb}
\Tmb=f \frac{h\nu}{k} (\frac{1}{e^{h\nu/kT_{\rm ex}}-1}-\frac{1}{e^{h\nu/kT_{\rm bg}}-1})(1-e^{-\tau}),
\end{equation}
where the background temperature $T_{\rm bg}=2.73 \K$ and $\nu$ is the line frequency.

We assume that the \twCO~\Jotz\ emission is optically thick, with 
$\tau_{\rm ^{12}CO}  \gg 1$, 
which is a valid approximation for MCs with \HH\ column density 
$\NHH \gtrsim 10^{21}~\cm^{-2}$ (see discussion in Section~\ref{sec:parameters}).
Under the assumption of local thermodynamic equilibrium (LTE), 
the excitation temperatures of different lines are the same and are equal to the kinetic temperature of the molecular gas $\Tk$. 
We obtain the kinetic temperature $\Tk$ ($=T_{\rm ex}$ in LTE) 
with the main-beam temperature of the \twCO~\Jotz\ emission:

\begin{equation}
\Tk=\frac{5.53}{\ln\{1+5.53/[\Tmb(^{12}{\rm CO})/f+0.84]\}} \K.
\label{equ:tk}
\end{equation}

\begin{figure*}[!htbp]
\epsscale{1.2}
\plotone{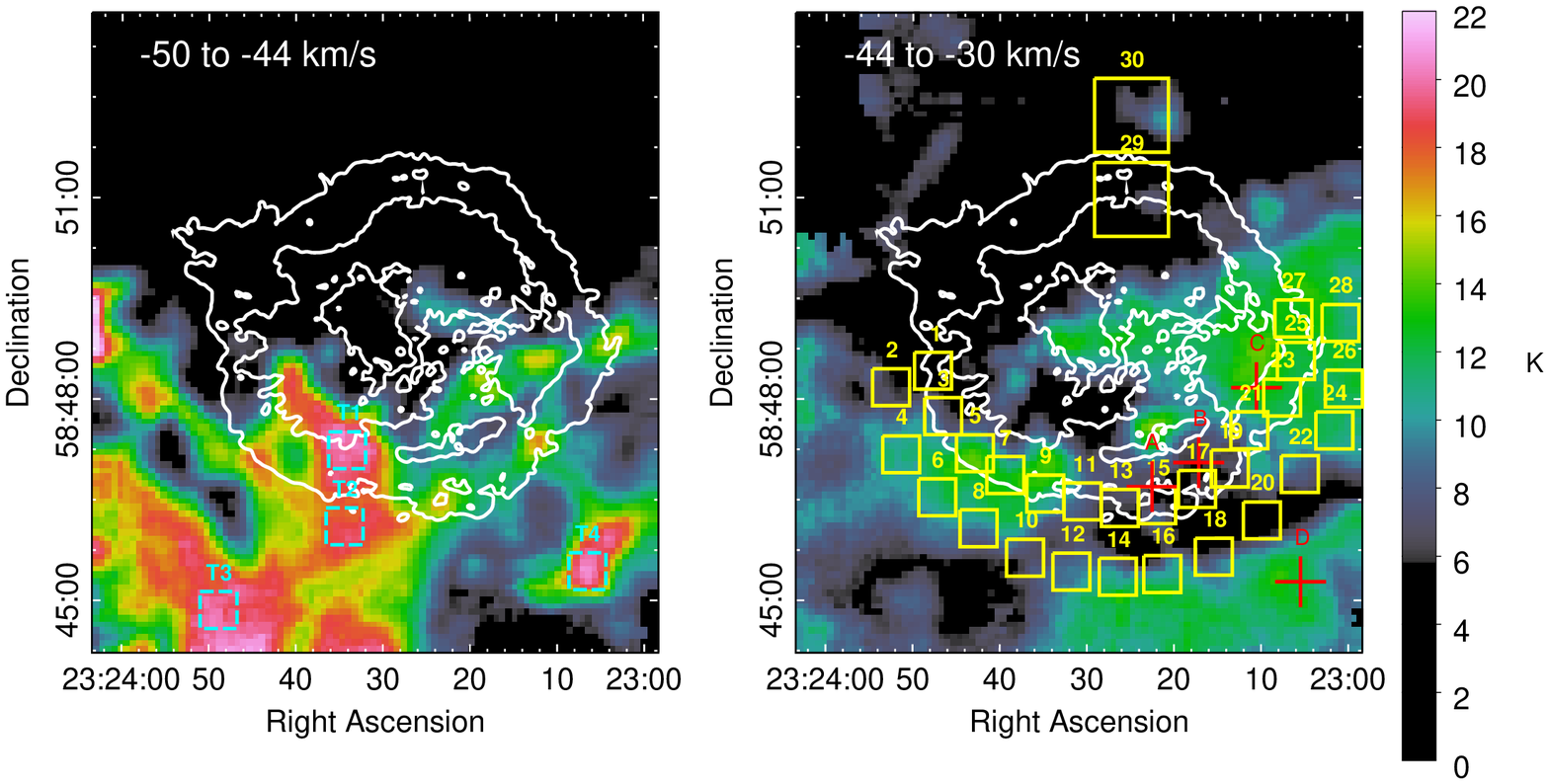}
\caption{
Kinetic temperature $T_{\rm k}$ maps of the molecular gas 
at velocities of $-50$ to $-44~\km\ps$ and $-44$ to $-30~\km\ps$,
respectively, assuming no beam dilution ($f=1$) and $\tau_{\rm ^{12}CO} \gg1$.
The real gas temperature is best represented by the 
high-temperature parts that are located in the MC interior, where the 
assumptions of $f=1$ and $\tau_{\rm ^{12}CO}\gg1$ are valid.
\twCO\ and \thCO\ spectra of the regions labeled with 
the boxes and plus signs are shown in Figure~\ref{fig:spec}.
\label{fig:Tmap}}
\end{figure*}

The gas temperature is thus inferred by the optically thick \twCO~\Jotz\ line with
given $f$.
Figure~\ref{fig:Tmap} shows $\Tk$ of the MCs at
two velocity intervals of $-50$ to $-44\km\ps$ and
$-44$ to $-30\km\ps$.
The latter velocity range is large and covers a few gas components
(see Figure~\ref{fig:12co21_grid}).
If multiple line components exist along the line of sight, 
the $\Tk$ is derived from the component with the highest $\Tmb$.
The $\Tk$ in each pixel is derived from $\Tmb$(\twCO) at
the given velocity range with an assumed beam filling factor $f=1$.
Note that the assumptions of $f=1$ and $\tau_{\rm ^{12}CO}\gg 1$ are not valid at 
the outer layers of  MCs, where $\Tmb$ is smaller owing to the clumpiness of the
gas and small optical depth. 
Therefore, the gas temperature at each velocity interval is 
best represented by the inner part of the clouds, where 
$f\approx 1$ and $\Tk$ appears to be the highest.

Notably, the MC at $-50$ to $-44$~\kms\ has a significantly 
higher temperature ($\sim 20~\K$) than all those 
in the velocity range of $-44$ to $-30$~\kms\ 
($\sim 13\K$, typical in interstellar MCs),
implying that an extra heating source is needed to warm 
the cloud.
The $20~\K$ warm gas is distributed both inside 
\snr\ (projected) and far outside the SNR, which does not support
the SNR shock being the direct heating source of the warm gas, otherwise,
we would only find hotter gas at/inside the SNR boundary.

\subsection{Molecular line profiles} \label{sec:line_profiles}
Physical interaction between SNR and MC involves
heating and perturbation of the MCs by the SNR shock.
The observational manifestations of the SNR-MC interaction 
in molecular lines include the presence of broadened,
optically 
thin lines with enhanced high-$J$ to low-$J$ ratios 
of molecular lines due to a high excitation
(e.g., \twCO\ \Jtto/\Jotz\ ratio $R_{12/10}
\approx 4 \exp (-11.1/T_{\rm k})>1$), etc.
See \cite{jiang10} and references therein for the SNRs showing 
these properties.
The heated clouds are found to have temperatures from tens of kelvin 
in the low-velocity C-type shock 
\citep[e.g.,][]{vandishoeck93} to hundreds of kelvin in 
the high-velocity J-type shock \citep{rho15}. 

\begin{figure*}
\epsscale{1.2}
\plotone{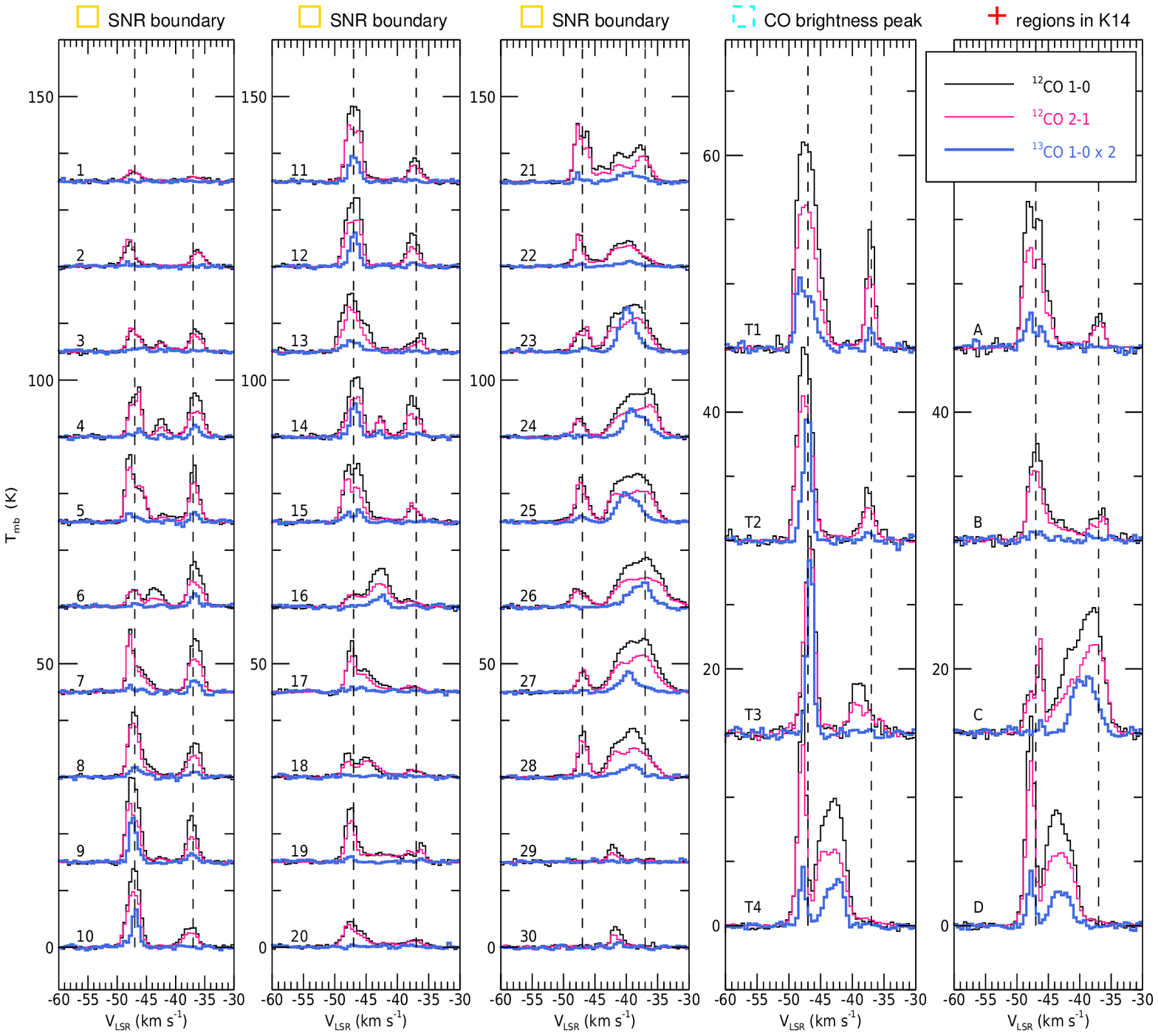}
\caption{
The \twCO\ \Jotz\ (black line), \twCO\ \Jtto\ (pink line), and \thCO\
\Jotz\ (blue thick line) spectra extracted from the regions
labeled in Figure~\ref{fig:Tmap}. Three groups of regions are  
selected: regions near the SNR boundary labeled with numbers 1--30 (columns 1--3;
odd and even numbers 
for regions inside and outside the SNR, respectively),
\twCO\ brightness peak labeled with T1--4 (column 4),
and  regions considered to have broad CO line width in Kilpatrick
et al.\ (2014),
labeled with A--D (column 5; spectra average from 
$\sim 50''\times 50''$ boxes).
The two vertical dashed lines indicate the velocities at
$-47\km\ps$ and $-37\km\ps$, respectively.
\label{fig:spec}}
\end{figure*}

To find out whether the MCs are perturbed or heated by \snr, 
we examine the line profiles of 
\twCO\ \Jtto, \twCO\ \Jotz, and \thCO\ \Jotz\ in three groups of regions:
(1) near the boundary of \snr, where the SNR--MC interaction might
occur;
(2) at the CO $\Tmb$ peaks (search for line broadenings due to shocks);
and (3) where slightly broad CO line widths ($>6\km\ps$) were found 
by \citet{kilpatrick14}.
The regions are labeled in Figure~\ref{fig:Tmap} with cyan boxes, blue dashed
boxes ($33''\times 33''$), and red plus signs, respectively.
The \twCO\ \Jtto\ datacube used here is
convolved to the beam size of the \twCO\ \Jotz\ datacube.
We search for broadened lines with FWHM
$\Delta V$ significantly larger than that of the  environmental gas.
The \twCO\ emission from the shocked gas is expected to 
be optically thin, characterized by larger
\twCO/\thCO\ line ratio \citep[close to an abundance ratio of \twCO\ to \thCO\ of $\sim 60$;][]{lucas98}. 
The shocked MCs should also be heated to higher temperatures, 
with $R_{21/10}>1$ and $\Tmb$ that differ from those in the 
preshock regions.
Note that the 20~K warm clouds outside the SNR also 
show $R_{21/10}  >1$ at the optically thin case,
but given their location, they cannot be heated by the SNR shock.
We must take the line widths into account, so as to 
distinguish the SNR-shocked gas and the gas with other extra heating.

First, we extract lines from the edge of \snr\ (denoted by
odd numbers), and compare them with those from nearby ``preshock'' regions 
(denoted by even numbers; see regions
in Figure~\ref{fig:Tmap} and line profiles in Figure~\ref{fig:spec}).
There are two main line
components ($\sim -47\km\ps$ and $\sim -37\km\ps$) in 
a small velocity range ($-50$ to $-30~\km\ps$), each with 
a velocity width of a few $\km\ps$.
None of the lines show $\Delta V \gtrsim 10  \km\ps$, which
is expected for shocked gas that has higher velocity dispersion
than the environmental gas with $\Delta V$ up to 6--7~\kms\ (regions 24, 26, 28).
Moreover, the line profiles at the SNR edge are not broader
than those from ``preshock'' regions.
A few regions (5, 7, 13, and 17) show narrow redshift \twCO\ wings 
of the $\VLSR=-47\km\ps$ component, which are indistinguishable 
from profiles of interstellar MCs that have 
line crowding along the line of sight 
(check external regions 16 and 18).
Although in some regions (2, 3, 5, 6) the $R_{21/10}$ is 
marginally larger than 1 at  $\VLSR\sim 47\km\ps$, 
the line widths of about $2\km\ps$ are too narrow to be considered shocked features.
In western regions 23--28,  the \thCO\ emission has 
$\Delta V \sim 3$--$4\km\ps$ at 
$\VLSR\sim -39\km\ps$. 
Despite the slightly larger line widths ($\sim 6\km\ps$), 
the \twCO\ lines reveal flatter tops, indicating that
the \twCO\ profiles could be affected by the opacity 
broadening \citep{phillips79,hacar16}.
Moreover, given that the \twCO\ emission is optically thick 
and has a small $\Tmb$ ($\lesssim 10~\K$), the western MCs 
consist of cold gas that has not been heated by the shock.
Therefore,  the line profiles in these regions do not reveal
evidence of shock--MC interaction.

Second, some clouds at $\VLSR \sim -47~\km\ps$ show strong
\twCO\ emission ($\Tmb\gtrsim 15~\K$; regions are labeled in 
Figure~\ref{fig:Tmap}). 
As shown in the fourth column of Figure~\ref{fig:spec}, the
\twCO\ and \thCO\ lines at $\sim -47~\km\ps$ are narrow 
and optically thick, which are properties of quiescent 
molecular gas.
The reason for the high $\Tmb$(\twCO) for these regions is that they are
at the interior parts of the MCs, with $\tau_{\rm ^{12}CO} 
\gg 1$ and filling factor $f\approx1$ (see Equation~(\ref{equ:tmb})).

\begin{figure*}
\epsscale{1.2}
\plotone{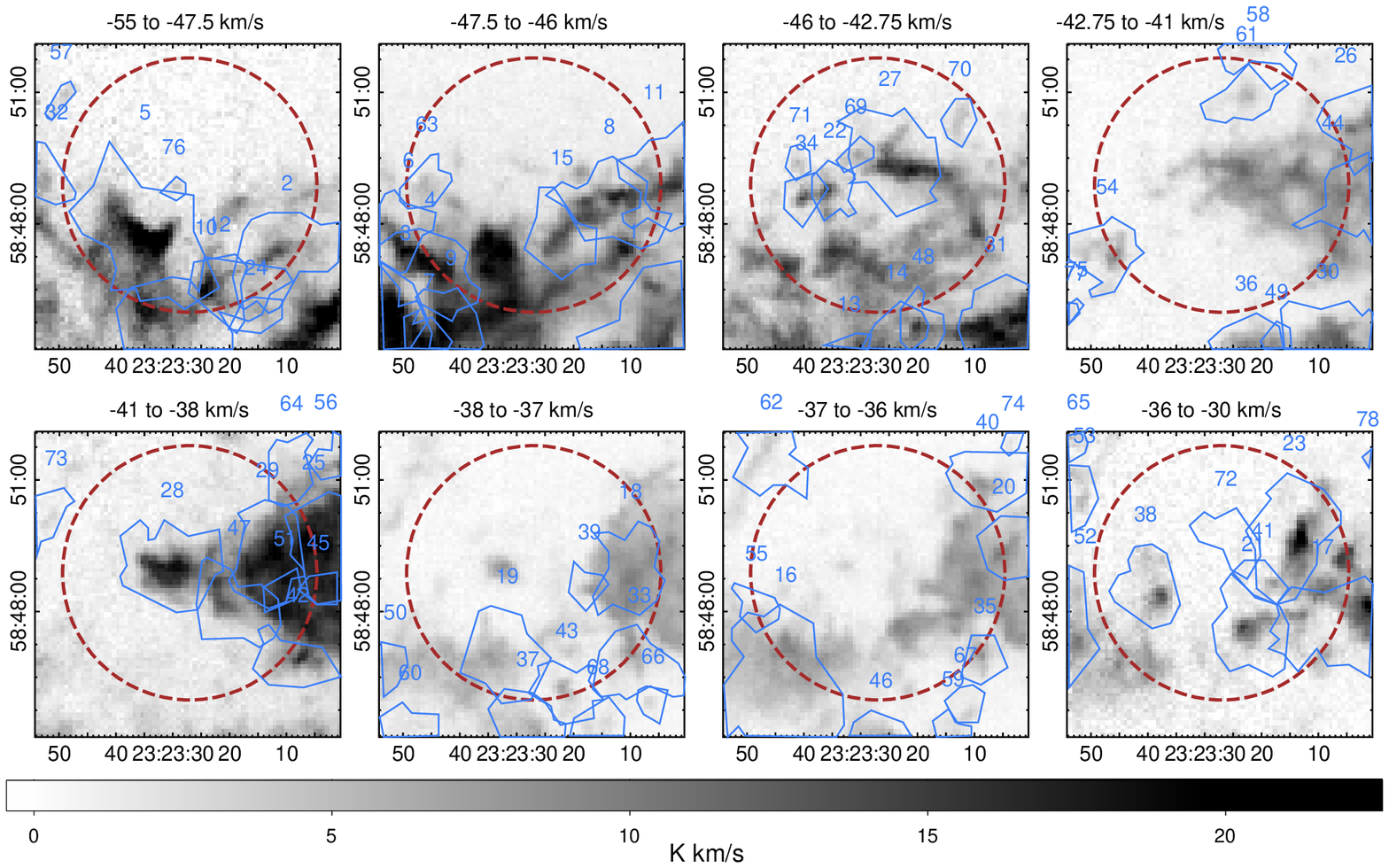}
\caption{
78 identified clumps overlaid the \twCO~\Jtto\ images at eight
different velocity ranges.
The first panel shows 8 clumps  with peak velocities within the 
velocity range labelled above, and for the rest panels 
each image shows 10 clumps.
The polygon region is a fit to the outer boundary of the clump.
The dashed circles indicate the the approximate shock
front of the SNR (radius of $2\farcm{9}$).
The detailed information of the clumps is tabulated in
Table~\ref{tab:12co21_clumpfind}.
\label{fig:12co21_clumpfind}}
\end{figure*}

Finally, we show the line profiles of four regions (A--D), 
where the broad \twCO\ \Jtto\ line width $\Delta v> 6\km\ps$ was suggested to
be due to the SNR shock \citep{kilpatrick14}.
We checked the \twCO\ lines and compared them with the \thCO\ lines 
(see also our Figure~\ref{fig:Tmap}).
The \twCO~\Jtto\ lines from regions A--C reveal similar 
profiles to those in \citet[][see their Figure 5]{kilpatrick14}, 
except that we found that the emission is twice as strong.
The \twCO\ line at region D shows a stronger line at 
$-47\km\ps$, which could be a result of the better 
angular resolution of IRAM 30~m data compared to 
the previous SMT data for resolving
the small molecular clumps.
By comparing the three lines, we notice
that all regions have $R_{21/10}\le 1$, 
and the slightly  ``broad''  \twCO\ lines in clouds 
A, B, and D are clearly optically thick. 
For the same reasons that we stated for regions 23--28, there 
is no strong evidence for shock--MC interactions in regions A--D 
as well.

\subsection{Clumps decomposed from the MCs} \label{sec:clumpfind}

The MCs are clumpy (see  Figure~\ref{fig:12co21_grid}) 
and the \twCO\ lines are found to be crowded in some regions.
Instead of showing the spectra of each pixel,
we decompose the MCs into multiple disjoint clump components,
each associated with a single significant \twCO\ emission peak. 
This would allow us to find whether there are any hot CO clumps
with high intensity and CO lines with broad widths.

Using the \twCO\ \Jtto\ datacube in $\VLSR=-50$ to $-30\km\ps$
with a velocity resolution of $0.25~\km\ps$,
we identified 78 clumps in a $7'\times 7'$ region 
centered at ($\RA{23}{23}{27}$, $\decl{58}{48}{40}$, J2000),
by applying the FellWalker clumpfind algorithm 
\citep{berry15} in the STARLINK package \footnote{http://starlink.eao.hawaii.edu/starlink}.
Figure~\ref{fig:12co21_clumpfind} displays the regions of the
clumps on the \twCO~\Jtto\ images at different velocity ranges. 
Each image shows clumps with peak velocities within the 
same velocity range.
The identification method and the information of the clumps 
are elaborated in the Appendix~\ref{sec:clump} and Table~\ref{tab:12co21_clumpfind}.

The \twCO\ \Jtto\ clumps have peak main-beam temperatures
$T_{\rm mb}^p$ 
in the range 2.2 -- 16.7~K, velocity dispersions $dv$ 
in the range 0.3--2.5~\kms, and sizes from subparsec to 3 pc.
Under a Gaussian assumption of the lines, the velocity 
widths $\Delta V$ of these clumps are about 0.7 -- $6 \km\ps$ (FWHM =
2.355$\sigma$).
The real line profiles could deviate from Gaussian distribution
for some of the clumps.
However, the small velocity dispersions do not favor  
the existence of bright, broad CO lines, consistent
with the results obtained from inspecting the line
profiles (see Section~\ref{sec:line_profiles}).

\subsection{Absorption line of \HCOp\ at $-50$ to $-33\km\ps$} \label{sec:hcop}

\begin{figure}
\epsscale{1.1}
\plotone{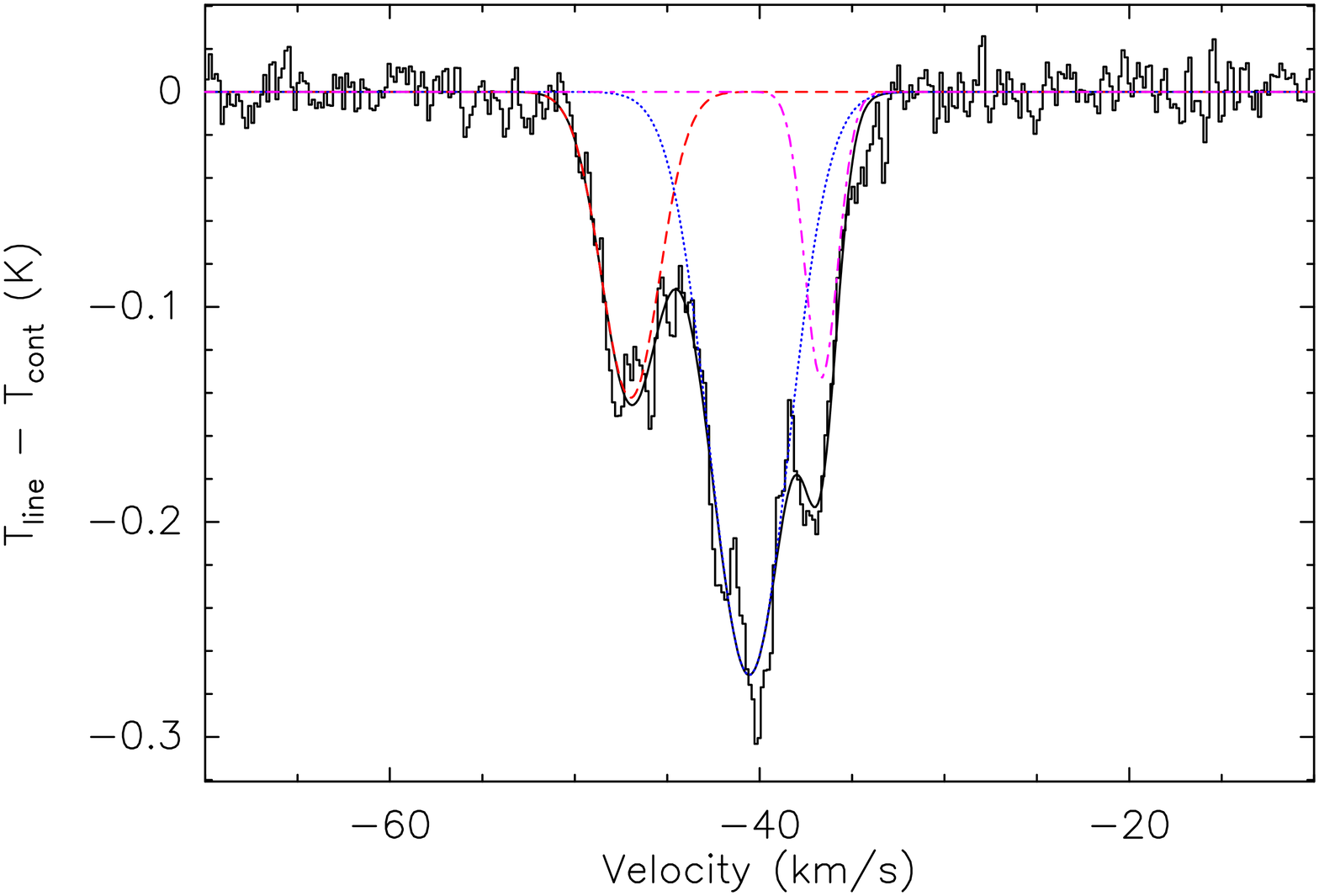}
\caption{
Continuum-subtracted spectrum of HCO$^+$ \Jotz\ at the radio peak of \snr\
($\RA{23}{23}{10}$, $\decl{58}{48}{40}$, J2000),
fitted with three Gaussian lines.
The sum of the three Gaussian components is shown with the black 
solid line. 
\label{fig:hcop}}
\end{figure}

We detected  absorption lines of HCO$^+$ \Jotz\
at around the radio peak in the west of \snr\
($\RA{23}{23}{10}$, $\decl{58}{48}{40}$, J2000; see
spectrum in Figure~\ref{fig:hcop}).
The absorption line at $-50$ to $-33\km\ps$ suggests 
that the background radio brightness exceeds the 
excitation temperature of HCO$^+$, and that
the HCO$^+$ gas at this velocity range 
is foreground gas. 

The column density of \HCOp\ can be obtained using the absorption
the spectrum and the given the brightness of the continuum.
The brightness temperature of the SNR's synchrotron emission 
at the radio peak is 1.85~K at the frequency of \HCOp\ \Jotz.
It is derived using the brightness temperature of 4.07~K at 32~GHz
(observed with Effelsberg; Kraus et al.\ in prep)
and a spectral index of $-0.77$ \citep{baars77}.
The brightness temperature of the
continuum is thus $T_{\rm cont}\approx 4.6~\K$ after adding the
contribution of the cosmic microwave background with $T_{\rm CMB}=2.73~\K$.
The column density of HCO$^+$ is given by \citep{godard10}

\begin{equation}
N({\rm HCO^+})=\frac{8.7\E{11} \int \tau(v) dv}{1-\exp(-h\nu/kT_{\rm ex})}
\cm^{-2}
\end{equation}
where the frequency $\nu=89.188$~GHz, 
the optical depth $\tau=-\ln[T_{\rm line}/T_{\rm cont}]$.

\begin{deluxetable}{lccc}
\tablecaption{Fit results of the HCO$^+$ spectrum near the radio peak \label{tab:hcop}}
\tablehead{
\colhead{$\VLSR$} & \colhead{FWHM} & \colhead{$T_{\rm line} -T_{\rm cont}$} 
& \colhead{$N$(HCO$^+$)\tablenotemark{a}}\\
\colhead{(\kms)}& (\kms) & (K) & (cm$^{-2}$)
}
\startdata
$-46.98\pm 0.06$ & $3.7\pm 0.1$ & $-0.14\pm 0.02$ &  $1.3\E{11}$ \\
$-40.56\pm 0.04$ & $5.1\pm 0.2$ & $-0.27\pm 0.02$ &  $3.6\E{11}$\\
$-36.66\pm 0.04$ & $2.1\pm 0.1$ & $-0.13\pm 0.02$ &  $7.1\E{10}$ \\
\enddata
\tablecomments{The 1$\sigma$ uncertainty is given.
\tablenotetext{a}{A lower limit of the 
\HCOp\ column density obtained by using $T_{\rm ex}
=2.73 \K$  and $T_{\rm cont}= 4.6$~K.
}
}
\end{deluxetable}

We use three Gaussian lines to fit the \HCOp\ spectrum and
show the fitted results in Table~\ref{tab:hcop} (also see
the Gaussian components in Figure~\ref{fig:hcop}).
The lower limits of the $N$(\HCOp) values  are tabulated, 
assuming that \HCOp\ is not collisionally excited, 
with an excitation temperature $T_{\rm ex}= 2.73$~K.
We obtain a total \HCOp\ column density of $5.7\E{11} \cm^{-2}$ in the velocity range $-50$ to $-33 \km\ps$.

\section{Discussion}
\subsection{Lack of evidence to support the SNR--MC interaction}

According to our low-$J$ \twCO\ and \thCO\ mapping of \snr, 
the upper limits of the line width and temperature
of the MCs are $\sim 7\km\ps$ and $\sim 22$~K, respectively.
The MCs along the line of sight do not show any of the 
properties that are typical for shocked MCs: 
(1) optically thin broad line with FWHM much larger than that of the environmental gas (e.g. $>10\km\ps$),
(2) broadened CO lines accompanied by $R_{21/10}>1$, 
and (3) MCs in post-shock regions much warmer than in 
the preshock regions.
Therefore, we suggest that there is no physical evidence
to support that the SNR is impacting the 
molecular gas. 
This is in accordance with the picture that \snr\ is still 
expanding within the progenitor's wind cavity, and the
preshock density is low as measured from X-ray 
observations \citep[$0.9\pm0.3\cm^{-3}$;][]{lee14}.

We note that a few regions along the SNR boundary 
(e.g., regions 5, 7, 13, and 17) show red-shifted 
\twCO\ line wings (with widths of a few km~s$^{-1}$; see Figure~\ref{fig:spec})
of the $\VLSR \sim −47 \km\ps$ lines.  
It is difficult, however, to ascribe them to 
the shock-MC interaction. 
First, the MCs in these regions are suggested to be in 
the foreground of the SNR (see Sections~\ref{sec:hcop} and \ref{sec:parameters}), 
which could not account for a disturbance 
showing a redshifted line feature. 
Secondarily, the small widths of the redshifted wings are not
consistent with the ram pressure balance of the 
cloud shock. The pressure balance can be formulated by
$\nHH v({\rm H_2})^2\sim n_0 v_{\rm f}^2$ \citep{mckee75}, where $\nHH$ and $n_0$ are
the densities in the cloud and 
the preshock intercloud medium, respectively, and $v({\rm H_2})$ and $v_f$ are the velocities of 
the cloud shock and the blast wave, respectively. 
For $n_0\sim 1~\cm^{-3}$ \citep{chevalier03, lee14} and $V_f=5200~\km\ps$ \citep{vink98a},
the cloud shock velocity would be 
$v{\rm (H_2)} \sim 1.6\E{2} [\nHH/10^3 \cm^{-3}]^{-1/2} \km\ps$ (MC density $\nHH\sim 10^3\cm^{-3}$; Section~\ref{sec:parameters}).
Actually, a molecular line with such a high velocity width is not detected 
in our observations.
A cloud shock with velocity $\gtrsim 50 \km\ps$ would easily dissociate ambient molecules \citep[e.g.,][]{hollenbach80}, but molecules may reform subsequently in post-shock regions and emit broad lines \citep{hollenbach89, wallstrom13}.
Therefore, these wing-like profiles are probably from quiescent
MCs along the line of sight with systemic velocities 
$\VLSR> -47~\km\ps$.

One should be aware that if the shocked layer 
is much thinner than the resolution of the telescope, 
the shocked features might be diluted and difficult
to observe. 
For example, a broadened CO line with $\Delta V=12 \km\ps$
was found near SNR Kes~79 with a resolution of 
$15''$, but such a broad line was not seen under a 
resolution of $\sim 1'$ \citep{zhou16b}.
In this case, one needs to consider other clues
for SNR--MC interaction, such as drastic enhancement 
of the post-shock density indicated by other 
wavelengths, which, however,
cannot be regarded as direct evidence.
Moreover, such clues have not been found in \snr.
Our CO~\Jtto\ observations, with the best angular
resolution so far, can resolve  a scale of 
0.18~pc ($11''$) at the distance of \snr.
According to our observations, the evidence of 
SNR--MC interaction is lacking.

Although we claim that there is no broad CO emission detected from interstellar MCs toward \snr, the remnant was
found to form CO molecules in the supernova ejecta behind the reverse shock \citep{rho09,rho12},
which emit very broad CO emission with $\Delta V$ of a few hundred $\km\ps$ \citep{wallstrom13}.
The velocities of these CO ejecta knots \citep[with $\VLSR$ down to $-5660\km\ps$,][]{rho12} are beyond the 
velocity range for us to search for shocked interstellar
MCs.

\subsection{Properties of the foreground MCs} \label{sec:parameters}

\begin{figure}
\epsscale{1.1}
\plotone{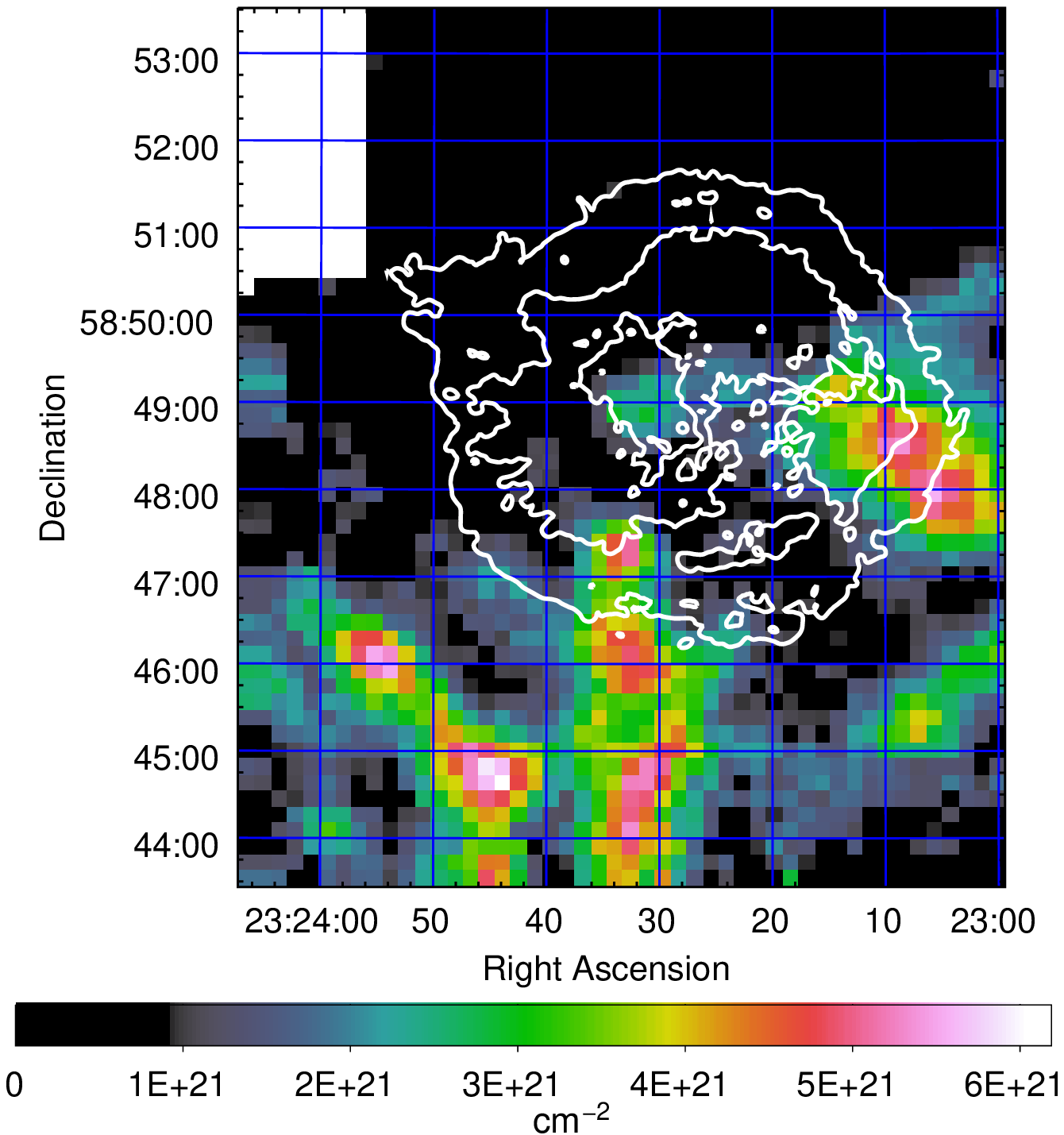}
\caption{
$\NHH$ distribution in the velocity range of 
$-50$ to $-30\km\ps$. 
\label{fig:nh2}}
\end{figure}

The MCs at $\VLSR=-50$ to $-30$~\kms\ are foreground gas,
as suggested by the \HCOp\ absorption line 
(see Section~\ref{sec:hcop} and Figure~\ref{fig:hcop}).
We derived the column density $\NHH$ and the molecular mass 
using \thCO\ lines, assuming that the lines are optically thin, 
with $T_{\rm k}=20~\K$ for the MC at $-50$ to $-44~\km\ps$ and 
$T_{\rm k}=13~\K$ for the MCs at $-44$ to $-30$~\kms.
Here we adopt the abundance ratio [\thCO]/[H$_2$] of $2\E{-6}$
\citep{dickman78}.
The foreground $\NHH$ distribution at $\VLSR=-50$ to $-30\km\ps$ is 
shown in Figure~\ref{fig:nh2}, with a maximum value of $\sim 6\E{21}~\cm^{-2}$ in the west and to the southeast of the SNR.
The MC component at $\VLSR\sim -2\km\ps$ is nearly 
uniformly distributed,
with a mean $\NHH$ of $1.6\E{20}~\cm^{-2}$ if it has 
$\Tk=10~\K$.
The total $\NHH$ in the foreground of \snr\ is, 
therefore, the sum of the two components.

The H$_2$ masses in the FOV are 1040~$d_3^2\Msun$ and $820~d_3^2\Msun$ for the velocity ranges of $-50$ to $-44\km\ps$ 
and $-44$ to $-30\km\ps$, respectively, where
$d_3$ is the distance of the MCs scaled to 3~kpc.
The 20~K warm MCs at $\VLSR= -47\km\ps$ 
contribute an H$_2$ mass of $\sim 200d_3^2 \Msun$ 
projected onto the SNR. 
This MC component can explain the detection of the 15--20~K dust 
reservoir of 2--4~$\Msun$  in front of the remnant
\citep{krause04, dunne03, barlow10},
given a dust-to-gas mass ratio of $\sim 0.01$. 
On the other hand, the cooler MCs in the velocity range 
$-44$ to $-30\km\ps$ contribute an 
H$_2$ mass of  $\sim 280d_3^2 \Msun$  projected onto the SNR.

The mean densities of the MCs are calculated by dividing 
the column densities by the cloud sizes.
At $\VLSR\sim -47 \km\ps$, the H$_2$ column densities at 
the \twCO\ emission peaks are $\sim2$--$6\E{21} \cm^{-2}$.
Given the cloud lengths of parsec scales (see Figure 
\ref{fig:12co21_clumpfind}), the cloud density is obtained as 
the order of $\nHH\sim 10^3 \cm^{-3}$.
At $-44$ to $-30~\km\ps$, the H$_2$ column densities of the MCs 
in the SNR west are $<5\E{21}~\cm^{-2}$. These clouds sometimes
overlap with others along the line of sight, with 
cloud lengths of parsec scales. 
Their typical densities are a few times $10^{2} \cm^{-3}$.
The H$_2$ densities estimated here are consistent with the 
values obtained in the previous CO observation \citep[$\le 4\E{3}~\cm^{-3}$;][]{wilson93} and the values of 
$\sim 3\E{2}$--$10^3 \cm^{-3}$ from the [CI] and 
\twCO\ study \citep{mookerjea06}.

\subsection{Foreground absorption of \snr}

The high foreground absorption of \snr\ is regarded as one
reason why the supernova explosion was likely not detected in
the seventeenth century \citep[e.g.,][]{reynoso02,krause08}.
Nevertheless, it is an open question whether the SN was 
witnessed by Flamsteed in 1680 at sxith magnitude \citep{hughes80}.
Recently, another possibility was proposed that
Cassini might discover the SN in or shortly before 1671 at fourth magnitude \citep{soria13}.
The foreground $\NHH$ at the explosion center of \snr\ \citep{thorstensen01} is  $\sim 2 \E{21}~\cm^{-2}$.
It gives a $V$-band extinction of $A_V \sim 2.0$ by applying
an $\NH$--$A_V$ conversion factor $\NH=2.87 \times10^{21} A_V$ cm$^{-2}$ \citep{foight16}
and   $\NH\approx 2.85\NHH$ \citep[the value of 2.85 accounts for a
photoabsorption cross-section ratio of H$_2$ to H;][]{wilms00}.
The extinction contributed by CO-traced
molecular gas is smaller than 
the $A_V$ derived from optical observations \citep[$=6.2\pm0.6$,][]{eriksen09},
implying that a lot of foreground materials
are in atomic phase or in the form of CO-dark gas 
\citep[e.g.,][]{oonk17,salas18}.

The MCs in front of \snr\ result in a heavier 
absorption of the SNR's emission in the west,
south, and center (see Figure~\ref{fig:nh2}). 
This is consistent with the fact
that larger absorption column densities
$\NH$ were found in these regions 
according to spatially resolved 
X-ray spectroscopy \citep{willingale02, yang08},
the study of OH lines \citep{keohane96},
and the extinction map based on {\it Herschel} observations
\citep{delooze17}.
The largest absorption caused by molecular gas 
in the west and south is 
$\NH\approx 2.85\NHH \sim 1.4\E{22}~\cm^{-2}$.
This $\NH$ value is around a half of that in 
\citet[$2.5\E{22}~\cm^{-2}$]{willingale02}, 
but similar to the result in \citet[$1.5\E{22}~\cm^{-2}$]{yang08}. 
The discrepancy of $\NH$ in two X-ray studies
could be due to the different atomic data or models
that they used.
The gas in molecular phase likely contributes $\sim 60\%$ 
of absorption in the west and south of the SNR and $\sim 40\%$ of the 
absorption near the SNR center, according to a comparison with the 
$\NH$ distribution obtained from the ISM dust \citep[see Figure F3 in]
[and the reference of $\NH/A_V$ conversion factor therein]{delooze17}.
Note again that we adopt the abundance ratio [\thCO]/[H$_2$] of $2\E{-6}$
\citep{dickman78} for calculating $\NHH$.

\subsection{The heating source of the warm MC at $\sim 47~\km\ps$}

The CO observation suggests that the SNR
suffers from heavier absorption by the MCs in the west,
the south and the center, with $\NHH$ up to $5\E{21}~\cm^{-2}$,
$5\E{21}~\cm^{-2}$, and $2\E{21}~\cm^{-3}$, respectively.

\begin{figure}
\epsscale{1.1}
\plotone{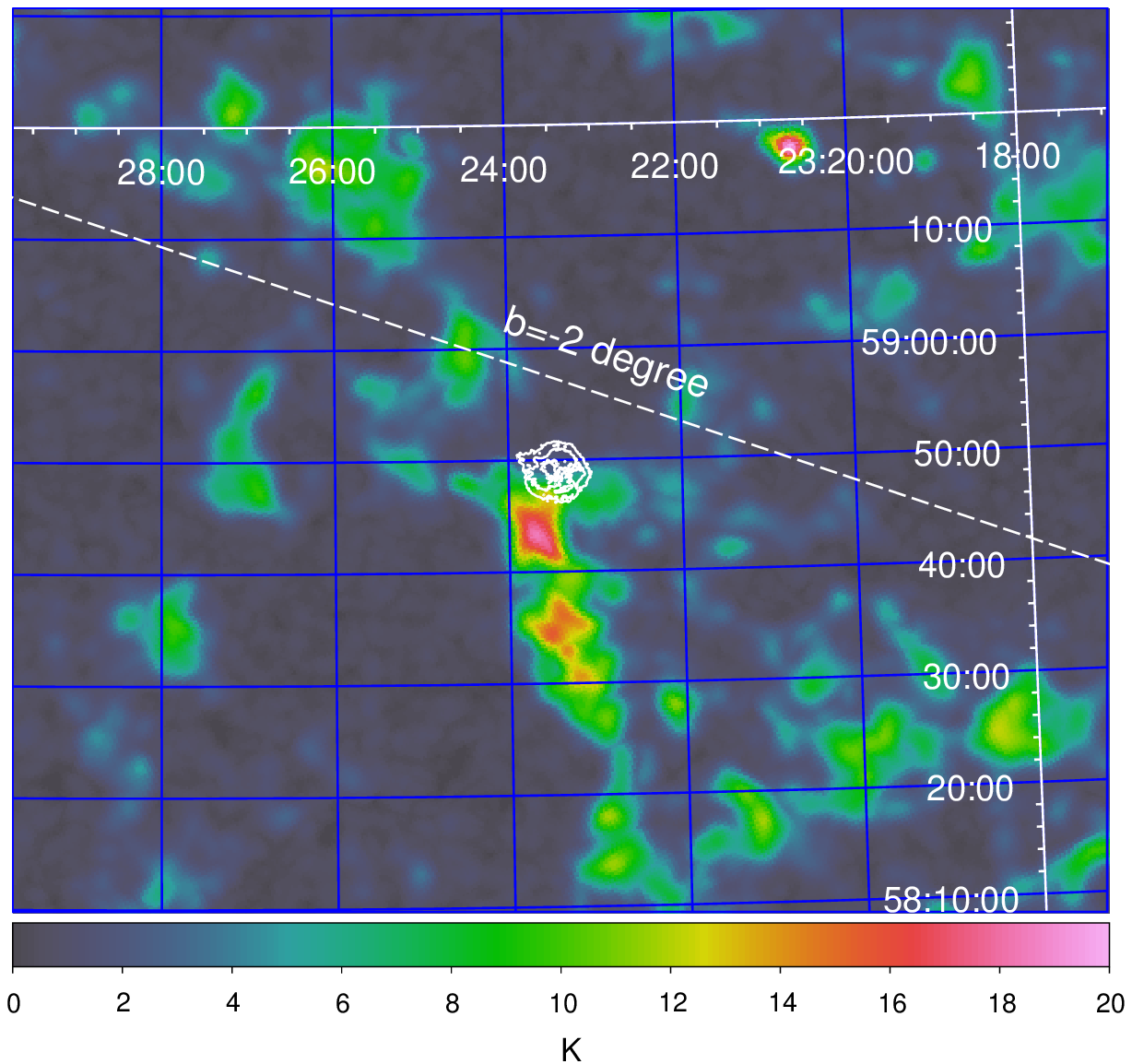}
\caption{
Large-scale FCRAO image of the  \twCO\ \Jotz\ main-beam 
temperature $\Tmb$ at $\VLSR=-50$ to $-44\km\ps$.
\label{fig:fcrao}}
\end{figure}

The foreground MCs at the velocity of $-47\km\ps$ have a 
temperature of around $20\K$, and narrow CO line widths of 
1--$3 \km\ps$ (see Figure~\ref{fig:spec}).
The clouds are warmer than those at other 
velocities (see Figure~\ref{fig:Tmap}).
Figure~\ref{fig:fcrao} shows a large-scale image of
the  \twCO\ \Jotz\ main-beam temperature $T_{\rm mb}$ in the velocity range of $-50$ to $-44\km\ps$
observed with FCRAO.
As indicated by the higher $\Tmb$,  the warmer gas 
is along a molecular ridge in the SNR south and extends 
over $12'$ ($10\du$~pc) south of \snr. 

We here discuss what heating sources 
cause the elevated temperature of the 20~K 
clouds.
The SNR shocks are not the direct heating source,
given that the clouds distribute both inside and 
far outside of the SNR.
Actually, the heating mechanism should be unrelated 
to any forms of recent turbulences or shocks, since 
the narrow CO line widths suggest that 
they are quiescent gas.
We can also exclude the progenitor winds of \snr\ as
the heating source, as there are no 
molecular line broadening or systematic shifts of 
line centers that may result from the winds.
The far-UV photons can ionize and heat 
neutral gas and dissociate molecules; however, 
they are typically heavily absorbed before reaching deep into the 
MCs with high column density ($>10^{21}~\cm^{-2}$).
For the far-UV-shielded regions in MCs,
the common heating mechanisms are
cosmic ray (CR) heating and X-ray heating.

\subsubsection{Background CRs}
Low-energy CRs ($< 1$~GeV) are 
the main  regulators for the ionization and chemical state of  dense cores lying deeply inside the far-UV-shielded 
regions.
CR-induced ionization is also accompanied by a heating 
to the molecular gas,  
with CR proton heating rate 
\citep{goldsmith78}

\begin{multline}
\Gamma_{\rm CR}=\zeta({\rm H_2})\Delta Q \nHH \\
\sim  3.2\E{-25}[\frac{\nHH}{10^3~\cm^{-3}}]
[\frac{\zeta({\rm H_2})}{10^{-17} \ps}] \erg \cm^{-3}\ps,
\end{multline}

where $\zeta({\rm H_2})$ is the CR ionization rate of H$_2$,
and $\Delta Q$ is the energy deposited as heat as a result
of the heating. A $\Delta Q$ of 20~eV is adopted here, but
it varies with CR energy \citep[17--26 eV for 10--100~MeV CR protons;][]{goldsmith78} and 
gas composition \citep[e.g.,\ 10--20~eV;][]{glassgold12}. 

The cooling of dense MCs is dominated by molecular 
lines and dust radiation.
For an MC with a density of $\sim 10^3~\cm^{-3}$, line cooling 
is approximated by \citep{goldsmith01}

\begin{equation}
\Lambda_{\rm gas}\sim 6\E{-24} (\frac{\Tk}{20~\K})^{2.4}
(\frac{dv/dr}{\km\ps \pc^{-1}}) 
\erg\cm^{-3}\ps
\end{equation}

where $dv/dr$ is the velocity gradient of the clouds,
which influences the optical depth $\tau \propto (dv/dr)^{-1}$. 

The molecular gas cooling via gas--dust interaction can be expressed as 
\citep{tielens05}

\begin{equation}
\Lambda_{\rm g-d}\sim 10^{-27} [\frac{\nHH}{10^3 \cm^{-3}}]^2 \Tk^{1/2} (\Tk-T_{\rm dust})\erg\cm^{-3}\ps
\end{equation}.

For the 20~K MCs with a density of $\sim 10^3~\cm^{-3}$, the molecular 
lines dominate the cooling. 
Therefore, we solve the kinetic temperature
from the thermal equilibrium $\Gamma_{\rm CR}=\Lambda_{\rm gas}$
for gas with $\nHH= 10^3~\cm^{-3}$, and obtain

\begin{equation}
\Tk\sim 20[\frac{\zeta({\rm H_2})}{1.9\E{-16}~\ps}]^{0.42} 
(\frac{dv/dr}{\km\ps \pc^{-1}})^{-0.42} \K
\end{equation}

CRs could heat the MCs to 20~K given an ionization 
rate $\zeta({\rm H_2})$ of $\sim 2\E{-16} \ps$ for 
a typical velocity gradient of $1\km\ps\pc^{-1}$.
\citet{neufeld17} obtained a CR ionization
rate of $2.8\E{-16} \ps$ per H atom for
the $-47\km\ps$ clouds toward \snr,
which was an updated value from the measurement
given by \citet[$\ge 2.5\E{-18}\ps$]{oonk17}.
Our estimated $\zeta({\rm H_2})$ level is 
reasonable among the range
measured recently in Galactic interstellar MCs
\citep[a mean value of $3.5^{+5.3}_{-3.0}\E{-16} \ps$; 
see][and references therein]{indriolo12},
but larger than the Galactic level suggested
in some earlier studies \citep[$10^{-17}$--$10^{-16}~\ps$; e.g.,][]{black77,vandertak00}.

The different $\Tk$ values 
between the $-47~\km\ps$ (20~K)
and $-44$ to $-30~\km\ps$ (13~K) can be explained 
even if they are exposed to the same level 
of $\zeta({\rm H_2})$.
The velocity dispersion of the western MCs at $\VLSR\sim -39\km\ps$ is about 3 times of those at 
$\VLSR\sim -47\km\ps$ (see Figure~\ref{fig:spec}), 
suggesting a larger velocity gradient 
$dv/dr$ and therefore stronger line cooling at the 
same density.

\subsubsection{CRs from \snr}
Could the CRs of \snr\ heat the MCs?
Previous observations have shown enhanced
CR ionization rate in the MCs close to SNRs W51C 
\citep[$\sim 10^{-15}~\ps$;][]{ceccarelli11}
and W28 \citep{vaupre14}. 
As CR ionization rate increases with CR 
energy density, the enhanced $\zeta({\rm H_2})$ 
near SNRs is indirect 
evidence that SNRs can accelerate CR protons.
However, it is difficult for the CRs of 
young SNRs to diffuse very far away from the
shock front.
The diffusion length of CRs from \snr\ over a
timescale of $t$ is \citep[see][and references therein]{vink12}
$l_{\rm diff} = \sqrt{2Dt}$, where the diffusion coefficient 
$D=\eta Ec/(3eB)$, with $\eta$ being the the deviation from 
Bohm diffusion.
We obtain $l_{\rm diff}=2.7\E{-3}\eta^{1/2} (\frac{E}{1~{\rm GeV}})^{1/2}
(\frac{B}{10 {\rm \mu G}})^{-1/2} (\frac{t}{340~\yr})^{1/2}~{\rm pc}$.
Only the very high energy CRs can reach further from the acceleration
site, and very high $\eta$ ($>100$) is required for a diffusion length as high
as 10~pc
(e.g., 100~TeV CRs can diffuse 8.7~pc way in the medium with $
B=10 \mu{\rm G}$ if $\eta=100$).
However, the diffusion coefficients in young SNRs cannot be so large, 
as indicated by the X-ray synchrotron emission well confined on the 
SNR shells  \citep[$\eta\lesssim 10$,][]{vink12}.

For the CRs from \snr, the proton energy above 10~TeV is 
$\sim 4.5 \E{47}\erg$, which is calculated using a proton 
spectrum with a power-law index of $-2.2$ and an exponential cutoff 
at 10~TeV and a proton energy of $5.1\E{48} \erg\ps$ above TeV \citep{ahnen17}.
These CR protons provide a mean CR energy 
density near \snr\ of  $u_{\rm CR} \sim 4.5\E{47} \erg/(4/3\pi 
l_{\rm diff}^3)\sim 3.7\E{-12}(l_{\rm diff}/10~\pc)^{-3}\erg\cm^{-3}$.
The thermal energy density needed by the MCs to raise a temperature
of $\Delta T=10~\K$ is  $\epsilon_{\rm MC}
=3/2 \nHH k\Delta T\sim 2\E{-12} [\nHH/10^3~\cm^{-3}]\erg\cm^{-3}$.
If the very high energy CRs from \snr\ were the heating
sources of the MCs, it would require the CRs to lose half of
their energy to MCs, which cannot be true.
First, the collision timescale between $>10$~TeV CRs and H$_2$ molecules is
$\tau_{\rm pp}=[2\nHH \sigma_{\rm pp} c]^{-1}\sim 5\E{3}\yr$, where
$\sigma_{\rm pp}\sim 10^{-25}~\cm^2$ is the 
cross section for proton--proton collision \citep{totem11} and 
$\nHH\sim 10^3~\cm^{-3}$ is used.
This means that the collision chance between the CR protons 
and nuclei in MCs is low for this 340 yr young SNR.
Only a very small fraction of high-energy CR protons can generate proton--proton 
collision in MCs and initiate  cascades of secondary particles.
Second, CRs more frequently transfer energy to the MCs through ionization 
rather than through proton--proton collision. 
The ionization timescale per CR particle in MCs is estimated to be 
$\tau_{\rm ion}=[\nHH \sigma_{\rm ion} c]^{-1}  \gtrsim 77 \yr$, 
where $\sigma_{\rm ion} < 1.4\E{-23}~\cm^2$ is the ionization cross section 
for $>10$~TeV CR protons \citep[derived from Eq.\ 31 in][]{rudd85}.
During the 340~yr, each CR proton may cause no more than 4.4 times of H$_2$ 
ionization on average while traveling back and forth across the shock front. 
Each ionization process only takes 37~eV from a CR proton to
generate an ion pair \citep{glassgold12}.
The mean total energy loss per CR proton is $\lesssim 160$~eV, which
is negligible in comparison with the CR energy of $>10$~TeV.  Hence,
even if $\eta$ has an extremely large value of $> 1000$
to allow the very high energy CRs to diffuse $>10$~pc away,
the CR energy is still insufficient to heat the MCs.

Therefore, the 20~K MCs at $-47~\km\ps$ cannot be heated by CRs from \snr, but by background CRs.

\subsubsection{X-rays from \snr}

\snr\ is a bright X-ray source with X-ray luminosity of $3\E{36} \erg\ps$
in 0.3--10~keV.
X-rays can penetrate large column densities 
of gas ($\NH>10^{21} \cm^{-2}$),
photoionize the MCs, and deposit part
of the energy to heat the MCs \citep{maloney96}.
For the X-ray emission with a photon index of 3.28, the X-ray energy 
heating rate per volume 
in 1--10~keV is \citep{maloney96}
\begin{multline} \label{Eq:gamma_x}
\Gamma_{\rm X}\sim 10^{-20}(\frac{\nH}{10^3 \cm^{-3}}) (\frac{F_{\rm X}}{\erg \cm^{-2} \ps})
(\frac{\NH^{\rm att}}{10^{22}~\cm^{-2}})^{-2/3} \\
\erg \cm^{-3} \ps
\end{multline}

where $\nH$ is the density of H nuclei,
$F_{\rm X}$ is the incident X-ray flux in 1--10~keV,
and $\NH^{\rm att}$ 
is the attenuating column density between the X-ray source
and the heated target. 
The atomic cloud envelope of the MCs contributes a column density $\NH\gtrsim 10^{22}~\cm^{-3}$ \citep[e.g.,][]{salas18},
which would have absorbed nearly all soft X-ray photons with energy 
$<1$~keV before the MC center is heated.
The fraction of energy transferred to heat is assumed to be 0.3.
Here a photon index of $3.28$  and a foreground 
absorption of $1.6\E{22}~\cm^{-2}$ are used to mimic the
X-ray spectral shape of Cas~A in 0.5--10 keV.
This simple model introduces $\sim 20\%$ uncertainties on the 
X-ray flux, since it does not fit the emission lines in the spectrum.

The youth of \snr\ must be taken into account for the
X-ray heating and molecular cooling.
The cooling timescale of the 20~K cloud is
slow, with $\tau_{\rm cool}=\epsilon_{\rm MC}/\Gamma_{\rm gas}\sim 2\E{4}\yr$.
It means that the MC cooling is not significant during the short lifetime of \snr\ and that the thermal equilibrium assumption cannot be used for this case.
On the other hand, the timescale of the incident X-ray photons absorbed by the 
${\rm H}_2$ gas is $\tau_{\rm abs}=[\nHH \sigma_{\rm H_2}c]^{-1}$,
where the photoabsorption cross section of H$_2$ is $\sigma_{\rm H_2}\sim 
2.85 \sigma_{\rm H}$ \citep{wilms00}, and 
$\sigma_{\rm H}(E)=2.6\E{-22} (E/ 1~{\rm keV})^{-8/3} \cm^2$ according to an empirical fit to
the experimental data in 0.5--7~keV \citep{maloney96}.
For X-ray photons with energies of 1 and 7~keV,  we obtain 
$\tau_{\rm abs}=1.4$~yr and 256~yr, respectively.
The soft X-ray emission dominates the heating, with an
absorption timescale much smaller than the SNR age.
Hence, it is only necessary to calculate heating rate based on the required thermal
energy.

To raise the MCs' temperature by 10~K (from a typical temperature 
of 10--20~K), the thermal energy required per volume is
$2\E{-12} [\nHH/10^3 \cm^{-3}] \erg \cm^{-3}$. 
If the thermal energy is constantly provided by the X-ray photons from \snr\ during
the past 340~yr,
the average X-ray heating rate is 
$\sim 2\E{-22}[\nHH/10^3 \cm^{-3}] \erg \cm^{-3}\ps$.
Comparing the value and Equation~(\ref{Eq:gamma_x}), the required
incident X-ray flux at the MCs is obtained as $F_{\rm X}\sim 10^{-2} (\NH^{\rm att}/10^{22} \cm^{-2})^{2/3} \erg\cm^{-3}\ps$.
Therefore, it requires the mean X-ray luminosity of \snr\ to be
$\sim 10^{38}(d_{\rm MC-SNR}/10~\pc)^2
\erg\ps$, where $d_{\rm MC-SNR}$ is the distance between 
the MC and the SNR, which should be $\gtrsim 10$~pc 
in order to explain the warm clouds south of \snr\
\citep[see also][for the suggestion of a larger distance of $>100$ pc]{kantharia98,salas17}.
Since the X-ray luminosity of a young SNR should increase along 
with the increasing swept-up  materials by the shocks,
the current X-ray luminosity of \snr\
is an upper limit for the SNR's lifetime.
Therefore, the required X-ray luminosity is over one
order of magnitude larger than the observed X-ray 
luminosity of \snr.
This suggests that the X-ray emission of \snr\ is insufficient to heat the MCs over 10~pc away.

\section{Conclusion}

We have performed mapping observations of \twCO~\Jotz, \twCO~\Jtto,
\thCO~\Jotz, and \thCO~\Jtto\ lines and \HCOp~\Jotz\ observation 
toward the SNR \snr\ with the IRAM~30 m telescope. 
Our main conclusions are summarized as follows.

\begin{enumerate}
\item 
\twCO, \thCO, and \HCOp\ emission is detected in the velocity
range of $-50$ to $-30\km\ps$, and at $\sim -2\km\ps$.
The MCs at $\VLSR=-47\km\ps$ have a kinetic temperature of 20~K,
narrow line widths of 1--$3 \km\ps$, column densities up to $6\E{21}~\cm^{-2}$,
and density of the order of $10^3\cm^{-3}$.
Some clouds are distributed in a wide velocity interval of
$-44$ to $-30\km\ps$, with a mean temperature of $13~\K$,
column density $<5\E{21}~\cm^{-2}$, and density of a few $\times 10^{2}~\cm^{-3}$.

\item
The MCs at $\VLSR=-50$ to $-30\km\ps$ are clumpy.
The identified \twCO\ \Jtto\ clumps have main-beam 
temperatures of 2.2 -- 16.7~K, velocity dispersions 
of 0.3--2.5~\kms, and sizes from subparsec to 3 pc.

\item 
The MCs toward \snr\ do not show any of the properties that are 
typical for shocked MCs: 
(1) optically thin broad line with FWHM much larger than that of
the environmental gas,
(2) broadened CO lines along with \twCO\ \Jtto\ to \Jotz ($R_{21/10})>1$, 
and (3) MCs in post-shock regions much hotter 
than in the preshock regions.
Therefore, we suggest that there is no physical evidence
to support that the SNR is impacting molecular gas.

\item 
We detect an absorption line of \HCOp\ \Jotz\ at $\VLSR=-50$ to $-33\km\ps$ 
near the radio peak west of the SNR, suggesting that all the detected molecular
gas is foreground gas.

\item 

The foreground MCs result in a high absorption of the SNR emission
at the west, south, and center of the SNR.
The 20~K warm gas contributes a foreground mass of 
$\sim 200 d_3^2~\Msun$  for \snr,
which can explain the cold dust (15--20~K; 2--4~$\Msun$) found in 
front of \snr.
The cooler MCs in the velocity range 
$-44$ to $-30\km\ps$ contribute a foreground  
H$_2$ mass of  $\sim 280d_3^2 \Msun$.

\item 
The gas at $\VLSR\sim -47\km\ps$ has a kinetic temperature of 20~K,
warmer than MCs at other velocities.
The warm gas is extended over 10~pc (projected distance) south of 
the SNR.
CRs are likely the heating source, and the required CR ionization rate
is $\zeta({\rm H_2})\sim 2\E{-16}\ps$. 
The CRs are provided by the local Galactic CR background, but not by
\snr, since the high-energy CRs from \snr\ do not provide enough energy
to heat the gas, and its low-energy CRs cannot diffuse far enough 
to the MCs.
The X-ray emission of \snr\ is insufficient to heat the MCs.

\end {enumerate}

\begin{acknowledgements}
We are thankful to Claudia Marka for the help on the IRAM observations. 
We also thank Alex Kraus for providing us with the Effelsberg 32~GHz image of \snr.
P.Z. acknowledges the support from the NWO Veni Fellowship, grant no. 639.041.647 and NSFC grants 11503008 and 11590781.
J.-T.L acknowledges the financial support from NASA through the grants 
NNX13AE87G, NNH14ZDA001N, and NNX15AM93G.
Y.C. acknowledges the support from the 973 Program grants 2017YFA0402600 and 2015CB857100 and NSFC grants 11773014, 11633007, and 11851305.
This work is based on observations carried out under project nos. 145-15, 053-16, 136-16, and 029-17 with the IRAM 30 m telescope. IRAM is supported by INSU/CNRS (France), MPG (Germany), and IGN (Spain).
The Canadian Galactic Plane Survey (CGPS) is a Canadian project with
international partners. The Dominion Radio Astrophysical Observatory
is operated as a national facility by the National Research Council
of Canada. The Five College Radio Astronomy Observatory CO Survey of
the Outer Galaxy was supported by NSF grant AST 94-20159. The CGPS is
supported by a grant from the Natural Sciences and Engineering
Research Council of Canada.
\end{acknowledgements}

\software{GILDAS/CLASS \citep{pety05},
Starlink \citep{currie14},
DS9,\footnote{http://ds9.si.edu/site/Home.html}
XSPEC (vers.\ 12.9.0u) \citep{arnaud96}.
}

\bibliography{snr}

\appendix

\section{clump decomposition} \label{sec:clump}

We apply the FellWalker clumpfind algorithm \citep{berry15}
in the STARLINK package 
to decompose the MCs to clumps. FellWalker is named as an analogy 
with the British pastime of ascending the hills of north England. 
It is a watershed algorithm that segments 2D/3D data by identifying the low-lying areas.
The maximum value of a clump is at the hill peak, while the
pixels on the route uphill but between the watershed and hill peak 
are assigned to this hill/clump. 
The minimum number of pixels (three dimensions) in each clump is 16. 
The detection limit of the clump signal is set to 2 rms (0.7~K). 

Table~\ref{tab:12co21_clumpfind} summarizes the detailed 
information of the identified clumps. 
The velocity dispersion is given by  $dv=(\Sigma T_i v_i^2/\Sigma T_i -(\Sigma T_i v_i/\Sigma T_i)^2)^{1/2}$, where $T_i$ is the 
intensity of the emission at pixel $i$ minus the background value,
and $v_i$ is the $\VLSR$ of pixel $i$.
For a clump with a Gaussian profile, the $dv$ is equal to the 
Gaussian standard deviation $\sigma$.

\input{12co21_fw_cat.tex}\label{tab:12co21_clumpfind}

\end{document}

%% file: 12co21_fw_cat.tex
\begin{center}
\begin{deluxetable*}{lccccc| lccccc}
\tabletypesize{\scriptsize}
\tablecaption{\twCO~\Jtto\ clumps decomposed from the MCs toward \snr}
\tablehead{
\colhead{clump} & R.A. & Decl. & $\Tmb^{\rm p}$ & $V_{\rm LSR}$ & $dV$  & 
\colhead{clump} & R.A. & Decl. & $\Tmb^{\rm p}$ & $V_{\rm LSR}$ & $dV$  \\
 & & & (K) & (\kms) & (\kms) & 
 & & & (K) & (\kms) & (\kms)
}
\startdata
 c1  &23:23:06.4 & 58:45:16.4
  &  16.7 & $-  47.5$ &   0.9
&
c40  &23:23:08.5 & 58:50:51.9
  &   7.0 & $-  36.5$ &   1.1
\\
 c2  &23:23:11.3 & 58:47:22.9
  &  14.6 & $-  47.5$ &   0.9
&
c41  &23:23:18.4 & 58:48:39.9
  &   6.8 & $-  35.8$ &   0.6
\\
 c3  &23:23:51.6 & 58:45:43.8
  &  14.5 & $-  46.8$ &   0.9
&
c42  &23:23:07.8 & 58:46:55.4
  &   6.8 & $-  41.0$ &   0.9
\\
 c4  &23:23:50.2 & 58:47:22.8
  &  14.4 & $-  46.0$ &   1.3
&
c43  &23:23:21.9 & 58:46:17.0
  &   6.6 & $-  37.5$ &   0.7
\\
 c5  &23:23:38.2 & 58:48: 1.4
  &  14.3 & $-  48.5$ &   1.4
&
c44  &23:23:01.4 & 58:48:28.8
  &   6.6 & $-  41.2$ &   0.9
\\
 c6  &23:23:50.9 & 58:47:39.3
  &  14.1 & $-  47.2$ &   0.9
&
c45  &23:23:00.7 & 58:48:28.8
  &   6.4 & $-  39.8$ &   0.4
\\
 c7  &23:23:46.7 & 58:45:10.9
  &  13.5 & $-  46.2$ &   0.9
&
c46  &23:23:26.2 & 58:45:11.0
  &   6.4 & $-  36.5$ &   0.5
\\
 c8  &23:23:11.3 & 58:48:39.9
  &  13.4 & $-  46.5$ &   0.8
&
c47  &23:23:21.9 & 58:48:45.5
  &   6.4 & $-  39.5$ &   1.4
\\
 c9  &23:23:43.1 & 58:45:32.9
  &  13.1 & $-  46.2$ &   0.9
&
c48  &23:23:19.8 & 58:45:38.4
  &   6.4 & $-  43.8$ &   1.1
\\
c10  &23:23:34.7 & 58:46: 0.4
  &  12.5 & $-  47.5$ &   1.3
&
c49  &23:23:19.1 & 58:45:10.9
  &   6.3 & $-  42.5$ &   0.6
\\
c11  &23:23:02.8 & 58:48:50.8
  &  12.1 & $-  47.0$ &   0.9
&
c50  &23:23:53.8 & 58:46:55.3
  &   6.2 & $-  37.8$ &   0.6
\\
c12  &23:23:24.1 & 58:46:28.0
  &  10.6 & $-  48.5$ &   0.7
&
c51  &23:23:09.2 & 58:48:39.9
  &   6.1 & $-  41.0$ &   1.0
\\
c13  &23:23:32.5 & 58:45:11.0
  &  10.4 & $-  46.0$ &   0.7
&
c52  &23:23:52.4 & 58:47: 0.8
  &   6.1 & $-  36.0$ &   0.8
\\
c14  &23:23:26.9 & 58:45:11.0
  &  10.3 & $-  46.0$ &   0.5
&
c53  &23:23:51.7 & 58:50:29.8
  &   5.9 & $-  35.5$ &   0.7
\\
c15  &23:23:21.2 & 58:48: 1.5
  &   9.8 & $-  47.2$ &   1.2
&
c54  &23:23:45.3 & 58:47:22.9
  &   5.8 & $-  41.8$ &   0.8
\\
c16  &23:23:42.4 & 58:47:11.9
  &   9.5 & $-  36.8$ &   1.0
&
c55  &23:23:48.1 & 58:47:44.8
  &   5.5 & $-  36.5$ &   0.6
\\
c17  &23:23:13.4 & 58:47:55.9
  &   8.8 & $-  36.0$ &   2.5
&
c56  &23:23:00.7 & 58:51:19.3
  &   5.2 & $-  39.8$ &   0.8
\\
c18  &23:23:10.6 & 58:48:39.9
  &   8.7 & $-  37.8$ &   1.4
&
c57  &23:23:48.9 & 58:50:57.3
  &   5.2 & $-  50.5$ &   0.3
\\
c19  &23:23:34.7 & 58:47: 0.9
  &   8.7 & $-  37.2$ &   0.8
&
c58  &23:23:20.5 & 58:52: 3.4
  &   5.0 & $-  41.2$ &   0.7
\\
c20  &23:23:05.6 & 58:49:18.3
  &   8.7 & $-  36.2$ &   1.1
&
c59  &23:23:12.7 & 58:45:10.9
  &   5.0 & $-  36.0$ &   0.5
\\
c21  &23:23:19.8 & 58:47:50.4
  &   8.5 & $-  36.0$ &   0.9
&
c60  &23:23:49.5 & 58:45:10.8
  &   4.6 & $-  38.0$ &   1.0
\\
c22  &23:23:36.1 & 58:48:45.4
  &   8.5 & $-  44.0$ &   1.1
&
c61  &23:23:22.6 & 58:50:52.0
  &   4.6 & $-  41.8$ &   0.7
\\
c23  &23:23:12.7 & 58:49:45.9
  &   8.5 & $-  34.8$ &   2.1
&
c62  &23:23:46.7 & 58:51:41.4
  &   4.3 & $-  36.2$ &   0.9
\\
c24  &23:23:14.2 & 58:46: 0.4
  &   8.0 & $-  47.8$ &   0.7
&
c63  &23:23:46.0 & 58:48:56.4
  &   4.3 & $-  47.2$ &   0.8
\\
c25  &23:23:02.8 & 58:49:23.8
  &   8.0 & $-  38.8$ &   1.4
&
c64  &23:23:09.2 & 58:50:46.4
  &   4.3 & $-  40.5$ &   0.7
\\
c26  &23:23:03.5 & 58:50: 7.8
  &   7.6 & $-  41.5$ &   1.0
&
c65  &23:23:53.8 & 58:51:57.8
  &   4.2 & $-  35.8$ &   0.7
\\
c27  &23:23:22.6 & 58:49:13.0
  &   7.5 & $-  43.5$ &   1.7
&
c66  &23:23:06.4 & 58:45:54.9
  &   4.0 & $-  37.8$ &   0.6
\\
c28  &23:23:34.0 & 58:49: 7.4
  &   7.4 & $-  39.2$ &   1.4
&
c67  &23:23:10.6 & 58:45:43.9
  &   4.0 & $-  36.5$ &   0.5
\\
c29  &23:23:11.3 & 58:49:18.4
  &   7.4 & $-  40.0$ &   1.6
&
c68  &23:23:16.3 & 58:45:10.9
  &   3.9 & $-  37.2$ &   0.6
\\
c30  &23:23:07.8 & 58:45:32.9
  &   7.4 & $-  42.0$ &   1.4
&
c69  &23:23:30.4 & 58:49:29.5
  &   3.7 & $-  45.8$ &   0.5
\\
c31  &23:23:06.4 & 58:45:27.4
  &   7.4 & $-  44.5$ &   1.0
&
c70  &23:23:12.0 & 58:50:24.4
  &   3.5 & $-  43.0$ &   0.5
\\
c32  &23:23:53.1 & 58:49: 7.3
  &   7.4 & $-  47.8$ &   0.5
&
c71  &23:23:41.0 & 58:49:23.9
  &   3.3 & $-  45.5$ &   0.6
\\
c33  &23:23:09.9 & 58:47: 6.4
  &   7.4 & $-  37.8$ &   1.0
&
c72  &23:23:24.1 & 58:49:13.0
  &   3.3 & $-  36.0$ &   0.5
\\
c34  &23:23:39.6 & 58:48:39.9
  &   7.3 & $-  43.5$ &   1.3
&
c73  &23:23:51.0 & 58:50:13.3
  &   3.2 & $-  38.2$ &   0.5
\\
c35  &23:23:08.5 & 58:47: 0.9
  &   7.3 & $-  36.2$ &   0.7
&
c74  &23:23:03.5 & 58:51:52.3
  &   2.8 & $-  37.0$ &   0.5
\\
c36  &23:23:23.3 & 58:45:11.0
  &   7.3 & $-  41.5$ &   1.1
&
c75  &23:23:53.8 & 58:46: 0.3
  &   2.7 & $-  42.8$ &   0.5
\\
c37  &23:23:26.9 & 58:45:16.5
  &   7.2 & $-  38.0$ &   0.7
&
c76  &23:23:29.0 & 58:48:51.0
  &   2.6 & $-  48.5$ &   0.5
\\
c38  &23:23:37.5 & 58:48:23.4
  &   7.2 & $-  35.0$ &   0.8
&
c77  &23:23:53.8 & 58:45:54.8
  &   2.3 & $-  42.2$ &   0.4
\\
c39  &23:23:16.3 & 58:48:28.9
  &   7.2 & $-  37.5$ &   0.8
&
c78  &23:23:00.6 & 58:51:30.3
  &   2.2 & $-  35.0$ &   0.3
\\
\enddata
\tablecomments{For each clump, the coordinates and $\VLSR$ are given for the pixel with peak \twCO\ \Jtto\ intensity $\Tmb^{\rm p}$, and $dv$ is the velocity dispersion of the clump.}
\end{deluxetable*}
\end{center}